\documentstyle[twocolumn,epsf]{jpsj}
\def \virg{\;\;,}
\def \point{\;\,.}
\def \kf{k_{\rm F}}
\def \vf{v_{\rm F}}
\def \e{{\rm e }}

\def \V2c{V_{\rm 2c}}

\def\ggs{\buildrel\textstyle > \over {\hbox{\raise0.2ex\hbox{$\sim$}}}}
\def\lls{\buildrel\textstyle < \over {\hbox{\raise0.2ex\hbox{$\sim$}}}}
\def\gsim{\,\lower0.75ex\hbox{$\ggs$}\,}
\def\lsim{\,\lower0.75ex\hbox{$\lls$}\,}
\def\runtitle{
 Competition of Dimerization and Charge Ordering 
  in  the Spin-Peierls State of  Organic Conductors
  }

\def\runauthor
{
 Muneo {\sc Sugiura}   
 and   
 Yoshikazu {\sc Suzumura} 
}

\title{ 
Competition of Dimerization and Charge Ordering 
 in the  Spin-Peierls State of  Organic Conductors  
}

\author{
 Muneo {\sc Sugiura} %$^{1,}$%
\footnote{E-mail: sugiura@slab.phys.nagoya-u.ac.jp} 
   and Yoshikazu  {\sc Suzumura} %$^{1,2,}$%
\footnote{E-mail: e43428a@nucc.cc.nagoya-u.ac.jp}
}

\inst{
Department of Physics, Nagoya University, Nagoya 464-8602 \\ 
}

\recdate{\hspace{35mm}}
\abst
{
The effect of the charge ordering on the spin-Peierls (SP) 
 state  has been examined 
  by  using a  Peierls-Hubbard model at quarter-filling
    with dimerization, on-site and 
      nearest-neighbor repulsive interactions.
By taking account of the presence of  dimerization,
  a  bond distortion  is calculated variationally 
 with the  renormalization group method based on  bosonization. 
  When   the charge ordering  appears at $V=V_c$
  with increasing the nearest-neighbor interaction ($V$),
    the distortion exhibits  a maximum   
       due  to  competition between 
       the dimerization and  the charge ordering. 
 It is  shown that the second-order phase transition  occurs 
    from  the  SP state  with the bond alternation 
      to a mixed state
      with an additional component of the site alternation 
     at  $V = V_c$. 
}

\kword
{
 spin-Peierls state, charge ordering, dimerization, 
 renormalization group, bosonization,  charge gap
  }

\begin{document}
\sloppy
\maketitle
%------ Introduction --------------------------------
\section{Introduction}
Recently  unconventional spin-Peierls (SP) states have been studied 
  extensively for  a system with a quarter-filled band, 
   which  shows a SP state  in the presence of a dimerization and/or 
    a  charge ordering (CO).
 Typical phenomena of such a state  have been   found in 
    quasi-one-dimensional organic conductors,  
    (TMTTF)$_2$X (X=PF$_6$,AsF$_6$),    
\cite{Bechgaard,Chow-CO} 
   which     undergo  the  SP transition  at temperature  
        being much lower than the CO temperature.
 The onset temperature of the CO state in the  TMTTF salts
\cite{Chow-CO} 
   has been identified  with the temperature corresponding to    
     the anomaly of the dielectric constant, i.e.,  
       the temperature at which 
       the   ferroelectric transition occurs.
\cite{Nad,Monceau}  
 In the several features of these  SP states, 
   the present paper concerns with  
    a coexistence and/or a competition of the SP state with  the CO 
 as  maintained in  the  following NMR experiment on (TMTTF)$_2$AsF$_6$.
\cite{Zamborszky}
 With increasing pressures up to 0.15 GPa, 
   the critical temperature of the SP state, $T_{\rm SP}$, increases 
     but that of the CO, $T_{\rm CO}$, decreases 
     where  the NMR line shapes 
      indicate  a coexistence of the  SP state and the CO state. 
 For  pressures larger than 0.15 GPa, 
   $T_{\rm SP}$ decreases gradually while the CO is absent.

Several SP states at quarter-filling  have been  studied theoretically  
   for  a  one-dimensional extended Hubbard model 
   coupled with a lattice
   by using  the variational calculation of the  distortion.
There are various  density waves.  
 The charge density wave (CDW) is the site-centered wave 
  with  a maximum on  the lattice site  
 and    the  bond order wave (BOW) is the bond-centered wave with  
  the  maximum  on the bond between  two neighboring lattice sites.
Two kinds of SP states have been obtained  
  depending on the location of the phase 
   of the  2$\kf$ wave and that of the 4$\kf$ wave
  where $\kf (=\pi/4a)$  is the Fermi wave vector 
  with $a$ being a lattice distance.
One of them is a state with 2$\kf$ BOW and 4$\kf$ BOW (state (a))
 and the other one is a state with 2$\kf$ CDW and 4$\kf$ CDW (state (b)) 
  where  4$\kf$ BOW and  
   4$\kf$ CDW represent  the dimerization and the CO, respectively.
The state (a) corresponds to  the state 
  with 2$\kf$ CDW$2$, 2$\kf$ BOW2 and 4$\kf$ BOW
     of  Ung {\it et al.}\cite{Ung},         
 the  $D_2$ phase of  Riera and Poilblanc, 
\cite{Riera}  and BCDW of Clay {\it et al.}\cite{Clay}, while 
 the state (b) corresponds to the state 
  with  2$\kf$ CDW$1$, 2$\kf$ BOW1 and 4$\kf$ CDW  of 
 Ung {\it et al.}\cite{Ung}, 
  and 4$\kf$ CDW-SP of Clay {\it et al.}\cite{Clay}
 Although the state (a) has been obtained explicitly 
  for  the large on-site  repulsive interaction,
\cite{Riera}
 it is complicated to find  the state (b). 
 The fact, that the CO state  relevant to the state (b)
  appears in the case of the large repulsive interaction between        
   electrons of the nearest-neighbor sites,
 has been shown   
 by      both  a numerical  diagonalization
\cite{Mila} 
  and  a mean-field theory.
\cite{Seo_Fukuyama}
   The  appearance of the CO state is understood  
   in terms of the phase Hamiltonian 
  where the CO is followed  by  the change  of the sign 
    of  the commensurate potential of the umklapp scattering  
     of the quarter-filled band.
\cite{Yoshioka_JPSJ00,Suzumura_JPSJ} 
There arises a competition between the CO state and 
 the SP state (i.e., the state (a)).
\cite{Clay,Sugiura2,Seo_Meeting}
 A phase diagram was obtained  as the functions   of 
  the electron-phonon couplings  for  
 intrasite and intersite displacements of the lattice, 
\cite{Clay}
 where the state (b) is obtained  only in the presence of 
 the intrasite  displacement. 
By calculating self-consistently the  state (a)
 in the presence of the dimerization but 
 without the state (b),
 the maximum of the bond distortion was obtained 
  at the onset of the CO.
\cite{Sugiura2} 
The state (b) for the extended Peierls-Hubbard model
 (i.e., only with the intersite electron-phonon coupling)
  has been first discovered 
 by Seo {\it et al.}
\cite{Seo_Meeting}
 who examined 
  the region of the CO state.
 Calculating  self-consistently  
    both the dimerization and the Peierls distortion,
   in which the phase of the  Peierls distortion is taken account,
  they obtained 
   a  first order transition between   
     the state (a) and the state (b),
    corresponding to  DM+SP and CO+SP respectively
      in their notations.
\cite{Seo_Meeting}
Thus the competition between the dimerization and the CO 
  is expected to exhibit a rich variety of the SP states.

 In the present paper, 
    treating the  dimerization as an external field, 
 we show the detail of the origin of  the maximum of $T_{\rm SP}$ 
 (the state (a))
 as a function of the nearest-neighbor repulsive  interaction 
  based on our  preliminary work.
\cite{Sugiura2} 
 Further we  
  investigate the  maximum of $T_{\rm SP}$   
    at the onset of the CO state   
      by considering 
     not only the state (b) but also 
     the states with the arbitrary phase of the Peierls distortion.
\cite{Seo_Meeting} 
The treatment of dimerization as the external field may  be reasonable for
   the study of the SP state of the organic conductors, TMTTF-salts, 
    since the dimerization does exist even at room temperatures
\cite{Ducasse} 
 and $T_{\rm CO}$ occurs at  much lower temperature.
In \S2, formulation is given. The Hamiltonian 
  is expressed in terms of the bosonization and 
  the renormalization group (RG)   equations are derived to 
  calculate the distortion, $u$, i.e., the order parameter of the SP state.
 In \S3, RG flows are calculated for both the state  without 
   CO  and that  with  CO. 
  The distortion $u$ is calculated as a function of the nearest-neighbor 
    repulsive interaction where a phase of the distortion is determined 
    to obtain the optimum $u$.  
 In \S4, discussion is given.

%------  Formulation  -------------------------------
\section{Formulation}
We consider a quarter-filled extended Peierls-Hubbard model  
with dimerization, given by
%------------------------(1)--------------------------------------
\begin{eqnarray}
 H  &=& - \displaystyle{ \sum_{j = 0}^{N-1} \sum_{\sigma}
        \left\{ t + (-1)^j x_d + (u_{j} - u_{j+1}) \right\} } 
        \nonumber \\
    & & \hspace{3.5cm} \times ( c^\dagger_{j,\sigma} c_{j+1,\sigma}  +  h.c.  ) 
        \nonumber \\
\nonumber \\
    & & + \, U \, \displaystyle{\sum_{j} n_{j,\uparrow}  n_{j,\downarrow}}
          + \, V \displaystyle{\sum_{j, \sigma , \sigma^\prime} n_{j, \sigma}  n_{j+1,\sigma^\prime}}
          \nonumber \\
    & & + \, \frac{K}{2} \, \displaystyle{\sum_{j} \left( u_{j} - u_{j+1} \right)^2} 
\virg 
\label{eq:Hamiltonian}
\end{eqnarray}
%-----------------------------------------------------------------------
where
%------------ (2) -----------------
\begin{eqnarray}
u_j = - u \cos(\pi j / 2 - \zeta ) \virg
\end{eqnarray}
 and
 $n_{j, \sigma} = c_{j, \sigma}^{\dagger} c_{j, \sigma}$.
 Quantities $u$ and $\zeta$ correspond to 
  an amplitude and a phase of  the Peierls 
 distortion, respectively and
  $c_{j, \sigma}^{\dagger}$ denotes 
   the creation operator for an electron
     with spin $\sigma$ at the site $j$. 
The $x_d$ term represents  the bond dimerization  while 
 $U$ and $V$ are coupling constants for  the  repulsive 
  interactions of  on-site and that of nearest-neighbor site, 
    respectively. 
The last term expresses the elastic energy induced by the distortion
  where $K$ is a spring constant.
 We use a coupling constant  $g (=4a/ \pi v_F K$) instead of $K$ ,
   where $v_F$ is the  Fermi velocity. 
 The  electron band splits  into an upper band 
 and a lower band due to  the dimerization $x_d$, where  
 the lower band is linearized  to study the state close to the Fermi 
  point. 
 By applying  the bosonization method, 
   we define phase variables $\theta_{\pm} (x)$ 
    and $\phi_{\pm} (x)$ as
%--------------   ( 3)   --------------------------------------------
\begin{eqnarray}
\theta_{\pm} \choose \phi_{\pm}
&=& \hspace{-0.2cm} \sum_{\sigma= \pm(\uparrow \downarrow) \atop q \ne 0} 
\hspace{-0.1cm} \frac{\pi {\rm i}}{q L} e^{- \alpha |q| / 2 - {\rm i} q x} 
\left[ \rho_{+,\sigma}(q) \pm \rho_{-,\sigma}(q) \right] 
 {1 \choose \sigma} \virg \nonumber \\
\label{eq:phase}
\end{eqnarray}
%--------------------------------------------------------------------
where the notation  $\sigma = +(-)$ denotes spin for  $\uparrow (\downarrow)$
 and  $\rho_{+(-),\sigma}$ 
 represents the density operator for the right going
 (left going) electrons.
The density operators $\rho_{+(-),\sigma}(q)$ 
 satisfy the boson commutation relation,
 $ [ \rho_{\pm,\sigma}(-q) \, \rho_{\pm,\sigma^\prime}(q^\prime) ] 
 = \pm (qL / 2 \pi) \delta_{q,q^\prime} \delta_{\sigma,\sigma^\prime}$.
In eq.~(\ref{eq:phase}), $\theta_+$ and $\phi_+$ represent the charge fluctuation  
 and the spin fluctuation, respectively,
since 
 $\pi^{-1} \partial \theta_{+} /\partial x$ 
($\pi^{-1} \partial \phi_+ /\partial x$) expresses 
 the charge density (spin density).
\cite{Suzumura} 
They satisfy the commutation relation,  
$
 [ \, \theta_{+} (x) ,  \theta_- (x^\prime) \, ] =  
   [ \, \phi_+ (x)  ,  \phi_- (x^\prime) \, ] =  
    {\rm i} \, \pi \, {\rm sgn}(x - x^\prime) \label{eq:k} \point 
$
Using phase variables, the operator for 
 the right going (left going) electron, $p=+(-)$, is written as
\cite{Luther_Peschel,Mattis,Suzumura} 
%----------(4)------------------------
\begin{eqnarray}
  \label{eq:Field_O}
\hspace{-0.6cm} \psi_{p,\sigma} (x)   
 &=&  \frac{1}{\sqrt{2\pi \alpha}}  
                    {\rm exp} \Big[{\rm i}  p  k_F  x  \nonumber \\
                  & & \hspace{-0.5cm} + \,  {\rm i} p \left[ \, \theta_+ + p \theta_- 
                             +   \sigma (\phi_+ + p \phi_-) \, \right] / 2 \Big] \,
                          {\rm e}^{ {\rm i} \Xi_{p,\sigma}} \, , 
\end{eqnarray}
%----------------------------------
where 
 $c_{j,\sigma} = \sqrt{a} \, ( \, \psi_{+,\sigma}(x) + \psi_{-,\sigma}(x) \, )$
  with  $x=aj$ and the lattice constant $a$. 
In eq.(\ref{eq:Field_O}), 
$ 
 \Xi_{+,\uparrow} = \pi (N_{+,\uparrow} + N_{-,\uparrow}) /2 
$,
$ 
 \Xi_{-,\uparrow} =- \pi (N_{+,\uparrow} + N_{-,\uparrow}) /2 
$,
$ 
 \Xi_{+,\downarrow} = \pi (N_{+,\uparrow} + N_{-,\uparrow})
   + \pi (N_{+,\downarrow} + N_{-,\downarrow}) /2
$ ,
$ 
 \Xi_{-,\downarrow} = \pi (N_{+,\uparrow} + N_{-,\uparrow})
   - \pi (N_{+,\downarrow} + N_{-,\downarrow}) /2
$ 
and 
$
 N_{p,\sigma} = \int dx \, \psi^{\dagger}_{p,\sigma} \psi_{p,\sigma}
$ .
In terms of these phase variables, 
eq.~(\ref{eq:Hamiltonian}) is rewritten as
\cite{Suzumura,Luther_Peschel,Mattis,Yoshioka_JPSJ00,Tsuchiizu},
%-----------------------------(5),(6),(7)----------------------------
\begin{eqnarray}
 H  &=& H_0 + H_1 \virg  
\label{eq:phase Hamiltonian} \\
 H_0  &=& 
      \; \frac{v_\rho}{4\pi} \int dx\;
  \left[\:\frac{1}{K_\rho} 
    (\partial_x \theta_+)^2+K_\rho (\partial_x \theta_-)^2\: \right] 
         \nonumber
                    \\
            & & \nonumber \\ 
  &+& 
      \;\frac{v_\sigma}{4\pi} \int dx\;
  \left[\:\frac{1}{K_\sigma} 
       (\partial_x \phi_+)^2+K_\sigma (\partial_x \phi_-)^2\: \right]
        \virg 
\label{eq:phase-Hamiltonian0}
  \\
            & & \nonumber \\
  H_1 &=& \; \frac{v_F}{2 \pi \alpha^2} \, 
         \int dx \, 
     \Bigl[ \, y_{1/4} \, \cos4\theta_+ \, 
                               \nonumber \\
  & &          - \, y_{1/2} \, \sin2\theta_+ 
                   + \, y_\sigma \, \cos2\phi_+ 
                                                    \nonumber \\ 
  & &  - (4\sqrt{2}\alpha u/\vf) \sin (\theta_+ + \zeta) \cos\phi_+
             \nonumber \\         
                    & & +         4 (\alpha u / \vf)^2 /g 
 \Bigr] \virg 
\label{eq:phase-Hamiltonian1}
\end{eqnarray}
%-----------------------------------
 where coefficients of the nonlinear terms  
 are given by 
%--------------  (8)   ---------------
\begin{eqnarray} 
  \label{eq:coupling}
  K_\rho   
      &=& \{ 1  / ( 1 + \tilde{U} + 4\tilde{V} ) \}^{1/2} ,
\nonumber \\
  K_\sigma  
       &=& \{ 1 / ( 1 - \tilde{U} ) \}^{1/2} ,  
\nonumber \\
 y_{1/4} &=&  A^4 (a / 2 \alpha )^2 \, 
           \tilde{U}^2 \, (\tilde{U}-4 \tilde{V}) ,
\nonumber \\
 y_{1/2} &=& 2(x_d /t) \tilde{U} / \{ 1 + (x_d /t)^2 \} , 
\nonumber \\
 y_{\sigma} &=& \tilde{U} \point
\end{eqnarray}
%-----------------------
The quantities $v_{\rho}$ and $v_{\sigma}$ are  velocities   given  by 
%-------------------------------------
$ 
 v_\rho = \vf (1 + \tilde{U} + 4\tilde{V} )^{1/2} 
$
and 
$
  v_\sigma = \vf ( 1 - \tilde{U} )^{1/2} 
$
where
$
 \vf = \sqrt{2} t a \{1 - (x_d / t)^2 \} / \{1 + (x_d / t)^2 \}^{1/2} ,
$ 
$
 \tilde{U} = U a / (\pi \vf) ,
$
$
 \tilde{V} = V a / (\pi \vf) 
$
and  $ A = \{ 1-(x_d/t)^2 \} / \{ 1+(x_d/t)^2 \} $.
The quantity $K_{\rho}$ ($K_{\sigma}$) expresses the degree of 
 the charge (spin) fluctuation.  
The amplitude, $y_{1/2}$, originates in  the umklapp scattering 
 of half-filling due to the dimerization 
while $y_{1/4}$  denotes that of quarter-filling 
obtained by integrating the 
 contribution from upper band \cite{Yoshioka_JPSJ00}.
The quantity $y_{\sigma}$ denotes the backward scattering. 
The cutoff parameter,
$
 \alpha
$, is of the order of the lattice constant
 and is taken as
$
 \alpha = 1.9 a / \pi
$
for quarter-filling.
\cite{Yoshioka_JPSJ00}
In terms of phase variables,
 the order parameters of 
$2 k_F$ CDW, 
 $4 k_F$ CDW and $4 k_F$ BOW 
 are written as  
%--------------------------(9),(10),(11)----------------------
\begin{eqnarray}  
O_{2 k_F \rm{CDW}} &=& \sum_{p,\sigma} \psi^\dagger_{p,\sigma} (x)
                                \psi_{-p,\sigma} (x) 
                       \nonumber \\
&=&  \frac{2}{\pi \alpha} \cos(2 k_F x + \theta_+ ) 
                           \cos\phi_+ \;, 
                   \label{eq:2kF-CDW}             \\
& & \nonumber \\
O_{4 k_F \rm{CDW}} &=& - \sum_p \psi^\dagger_{p,\uparrow} (x)
                         \psi^\dagger_{p,\downarrow} (x)
          \psi_{-p,\downarrow} (x) \psi_{-p,\uparrow} (x) \nonumber \\
      &=& - \frac{1}{2 \pi^2 \alpha^2} \cos(4 k_F x + 2 \theta_+) 
             \label{eq:4kF-CDW}     \virg  \\                         
& & \nonumber \\
O_{4 k_F \rm{BOW}}
   &=& - \sum_p \psi^\dagger_{p,\uparrow} \left( x-\frac{a}{2} \right) 
     \psi^\dagger_{p,\downarrow} \left(x+\frac{a}{2} \right) \nonumber \\
 & & \hspace{0.5cm} \times  \psi_{-p,\downarrow} \left(x-\frac{a}{2} 
              \right) \
        \psi_{-p,\uparrow} \left(x+\frac{a}{2} \right) \nonumber \\
&=&  \frac{1}{2 \pi^2 \alpha^2} \sin\left(4 k_F \left(x-\frac{a}{2} \right) + 2 \theta_+ \right) \point                          
             \label{eq:4kF-BOW}
\end{eqnarray}
%------------------------
 The variable  $x$ of  eqs.~(\ref{eq:2kF-CDW}) and 
   (\ref{eq:4kF-CDW}),
 is defined at  the lattice site 
  while that of  eq.~(\ref{eq:4kF-BOW}) is defined on  
   the bond,
   i.e. $x=a/2,3a/2, \cdot \cdot \cdot$ .
 Equations ~(\ref{eq:2kF-CDW}) and (\ref{eq:4kF-BOW})
  for  $\theta_+ \rightarrow   \pi/4$
 show  
  the 2$\kf$ BOW  and 4$\kf$ BOW while 
  eqs.~ (\ref{eq:2kF-CDW}) and (\ref{eq:4kF-CDW})
  for  $\theta_+ \rightarrow   \pi/2$ represent 
   2$\kf$ CDW and  4$\kf$ CDW.

 Quantities $u$ and $\zeta$ in eq.~(\ref{eq:phase-Hamiltonian1})
 are determined so as to minimize $\langle H \rangle$  leading to 
 two kinds of conditions , 
 $\langle \partial H / \partial u \rangle = 0$ and 
   $\langle \partial H / \partial \zeta \rangle = 0$. 
 The condition,  $\langle \partial H / \partial u \rangle = 0$, is written as
%-------------------------(12)-------------------------------
\begin{eqnarray}
 \hspace{-0.5cm} & & \frac{\sqrt{2} \alpha}{g v_F} u 
   =   F   \equiv 
     \big< \sin(\theta_+ + \zeta)\cos\phi_+ \big> 
                                               \nonumber \\
\hspace{-0.5cm} & =  & 
           \cos\zeta \, \big< \sin\theta_+ \cos\phi_+ \big> 
         + \sin\zeta \, \big< \cos\theta_+ \cos\phi_+ \big> 
  \, \virg 
\label{eq:SCE-p}
\end{eqnarray}
where $F$ is calculated as the function of $u$ and $\zeta$ 
 from eq.~(\ref{eq:phase Hamiltonian}). 
Instead of  the  condition of $\langle \partial H / \partial \zeta \rangle = 0$, 
 we use a condition of  
  optimizing  $u$ with respect to $\zeta$, i.e., 
 ${\rm d} u / {\rm d}\zeta = 0$, which is discussed later. 
 It is noted that  $F$ 
  is expressed in terms of the original Hamiltonian as  
%-----------------------------(13)---------------------------
\begin{eqnarray}
 F  &=&
 - \frac{\pi}{2 N}  \sum_{j, \sigma}  
   \left<  \sin \left( \frac{\pi j}{2}  - \zeta + \frac{\pi}{4} \right) 
   \right. \nonumber \\
& & \hspace{2cm} \left.  \times ( c^\dagger_{j,\sigma} c_{j+1,\sigma}  +  h.c.  )
   \right>  \point  
\label{eq:SCE}
\end{eqnarray}
%------------------------

The quantity   $u$ is  determined to satisfy the relation,
 $(\sqrt{2} \alpha / g v_F) u = F$.
The quantity $F$  of the r.h.s. of  eq.~(\ref{eq:SCE-p}) 
 is evaluated  by making use of the RG method, in which $u$ is 
  taken into account as the initial value.
In order to derive the RG equations, 
eq.~(\ref{eq:phase Hamiltonian}) is replaced by 
  an  effective Hamiltonian , 
 $H^{\rm{eff}} = H_0 + H^{\rm{eff}}_1$,  where 
 $H^{\rm{eff}}_1$ is written as 
%--------------------------------(14)--------------------------
\begin{eqnarray}
H^{\rm{eff}}_1   &=& \; \frac{v_F}{2 \pi \alpha^2} \, 
         \int dx \, 
     \Bigl[ \, y_{1/4} \, \cos4\theta_+ \, 
            - \, y_{1/2} \, \sin2\theta_+  \nonumber \\
   & & - \, y_{2 \rho}  \, \cos2\theta_+ 
     + \, y_\sigma \, \cos2\phi_+ \,  \nonumber \\
    & &
       - \, y_{ps} \, \sin\theta_ + \cos\phi_+ 
        - \, y_{pc} \, \cos\theta_ + \cos\phi_+ 
                                  \nonumber \\ 
    & &        +  \, (y_{ps}^2 + y_{pc}^2) /8g \, \Bigr] \virg 
\label{eq:phase Hamiltonian-RG}
\end{eqnarray}
%---------------------------- (15)  --------------------------------
 where  
\begin{eqnarray}
\label{eq:yps_ypc} 
 y_{ps} &=& ( 4 \sqrt{2} \alpha u / \vf )\cos\zeta  \virg 
\nonumber \\
 y_{pc} &=& ( 4 \sqrt{2} \alpha u / \vf )\sin\zeta   \virg
\end{eqnarray}
%-----------------------
and the $y_{2\rho}$ term  
     is  induced through the RG process
 with  increasing  the length of the scale, i.e.,   
   $\alpha \e^{l}$. 
 By applying a transformation, 
 $\alpha \to \alpha ( 1 + dl )$,
  to coefficients of eq.~(\ref{eq:phase Hamiltonian-RG}),
   \cite{Solyom,Giamarchi,Tsuchiizu,Yonemitsu}
    one obtains RG equations as
(Appendix)
%------------------------------(16a)  (16h) ------------------------
\begin{subequations}
\begin{eqnarray}
\label{eq:RGa} 
\hspace{-1.2cm} \frac{d}{dl} \, K_\rho(l)  &=& 
    - \, \Bigl[ \, 2\,y_{1/4} ^2(l) \,      
    + \frac{1}{2} \left( y_{1/2} ^2(l) + y_{2\rho} ^2(l) \right) 
       \nonumber \\  
& & \hspace{-0.3cm} + \, \frac{1}{16} \, \left( \, y_{ps} ^2(l) 
+ y_{pc} ^2(l) \, \right) \Bigr] 
      \, K_\rho ^2(l)  \virg 
\end{eqnarray}
\begin{eqnarray}
\label{eq:RGb}
\hspace{-1.5cm} \frac{d}{dl}\:G_\sigma(l)  &=&  
 - \, \; y_\sigma ^2(l) \;
 - \,\frac{1}{8} \, \left( \, y_{ps} ^2(l) + y_{pc} ^2(l) \, \right) 
             \,  \virg  
\end{eqnarray}
\begin{eqnarray}
\label{eq:RGc}
\hspace{-4.2cm} \frac{d}{dl}\:y_{1/4}(l)  &=&  
 \big[ \, 2 \, - \, 8 \, K_\rho(l) \, \big] \, y_{1/4}(l) 
                   \nonumber \\
& & \hspace{1.0cm}
 + \; \frac{1}{4} \left( \, y_{1/2}^2(l) - y_{2\rho}^2(l) \, \right) 
   \virg     
\end{eqnarray}
\begin{eqnarray}
 \label{eq:RGd}
\hspace{-1.cm} \frac{d}{dl}\:y_{1/2}(l)  &=&  
 \Big[ \, 2 \, - \, 2 \, K_\rho(l) \, + \; \frac{1}{2} \, y_{1/4} (l) \, \Big] \, y_{1/2}(l)\; 
\nonumber \\ 
  & & \hspace{-0.3cm} + \; \frac{1}{4} \, y_{ps}(l) \, y_{pc}(l)
          \virg 
\end{eqnarray}
\begin{eqnarray}
\label{eq:RGe}                  
\hspace{-1.cm} \frac{d}{dl}\:y_{2\rho}(l)  &=&  
 \Big[ \, 2 \, - \, 2 \, K_\rho(l) \, 
 - \; \frac{1}{2} \, y_{1/4} (l) \, \Big] \, y_{2\rho}(l)\;  \nonumber \\ 
& & \hspace{-0.3cm} 
 - \; \frac{1}{8} \, \left( \, y_{ps}^2(l) - y_{pc}^2(l) \, \right)
                   \virg  
\end{eqnarray}
\begin{eqnarray}
\label{eq:RGf}
\hspace{-0.5cm} \frac{d}{dl}\:y_\sigma(l)  &=& 
 - \, G_\sigma(l) \, y_\sigma(l)
 - \;\frac{1}{8} \, \left( \, y_{ps} ^2(l) + y_{pc} ^2(l) \, \right) \virg
\end{eqnarray}
\begin{eqnarray}
\label{eq:RGg}
\hspace{-1.cm} \frac{d}{dl}\:y_{ps}(l)  &=&  
 \Bigl[ \; \frac{3}{2} \, - \, \frac{1}{2} \, K_\rho(l) \, 
 - \, \frac{1}{4} G_\sigma(l) \, - \, \frac{1}{2} \, y_{2\rho}(l)  \nonumber \\ 
& & - \, \frac{1}{2} y_\sigma(l) \, \Bigr] \, y_{ps}(l) + \; \frac{1}{2}\;y_{1/2}(l) \, y_{pc}(l) \virg
\end{eqnarray}
\begin{eqnarray}
\label{eq:RGh}
\hspace{-1.cm} \frac{d}{dl}\:y_{pc}(l)  &=&  
 \Bigl[ \; \frac{3}{2} \, - \, \frac{1}{2} K_\rho(l) \, 
 - \, \frac{1}{4} G_\sigma(l) \, + \, \frac{1}{2} \, y_{2\rho}(l)  \nonumber \\ 
& & - \, \frac{1}{2} y_\sigma(l) \, \Bigr] \,y_{pc}(l) + \; \frac{1}{2}\;y_{1/2}(l) \, y_{ps}(l) \virg
\end{eqnarray}
\label{eq:RG}
\end{subequations}
%---------------------------------------------------------------------
where 
$
 K_\sigma = 1 + G_\sigma / 2 
$
and 
$
 G_\sigma = \tilde{U} 
$.
The initial conditions ($l=0$) for these quantities are given by 
 eqs.~(\ref{eq:coupling}) and (\ref{eq:yps_ypc}), 
 and $y_{2\rho}(0)=0$.
In deriving eqs.~(\ref{eq:RG}), 
 $v_{\rho}$ and $v_{\sigma}$ are replaced by $\vf$. 
We note that RG equations similar 
 to eqs.~(\ref{eq:RG}) have been derived by Yonemitsu,
\cite{Yonemitsu}
 who    treated the case of the phonon with finite frequency but 
    without  $y_{2\rho}$ and  $y_{pc}$ terms. 
Since eqs.~(\ref{eq:RG}) are the result of first order RG, 
 the renormalized quantity increases to infinity
  for the case of the strong coupling. 
Then when the renormalized quantity becomes large  
 ( e.g.,   $y_{1/2}, y_{2\rho},| y_{\sigma}|$, 
  $[y_{ps}^2+y_{pc}^2]^{1/2}$ 
 $\rightarrow$  2  in the present calculation ), 
  we stop the renormalization for  the corresponding quantity.

Here we estimate the quantity 
$F$ of eq.~(\ref{eq:SCE-p}) by employing the following response function,
  whose limiting value with  the long distance  gives  
    $F$   as  found   in the ordered state.       
 The  response function for $F$, which depends on both 
    the imaginary time $\tau$ and the space $x$, 
 is defined by 
%----------------------------(17)---------------------------
\begin{eqnarray}
\hspace{-0.7cm}  R(|\vec{r}_1 - \vec{r}_2|) &=& 
   \left< T_\tau \hat{F}(\vec{r}_1)  \hat{F}(\vec{r}_2) \right> \nonumber \\
&=& R_s \cos^2\zeta + R_{sc} \sin2\zeta + R_c \sin^2\zeta  \virg 
\label{eq:res}
\end{eqnarray}
%---------------------------
where
$
 \hat{F}(\vec{r})
 =    \cos\zeta \sin\theta_+ (\vec{r}) \cos\phi_+ (\vec{r}) 
         + \sin\zeta \cos\theta_+ (\vec{r}) \\
         \cos\phi_+ (\vec{r})  
$ 
  and $\vec{r}=(x,v_F \tau)$.
Quantities $R_s,R_c$ and $R_{sc}$ are given by
%--------------------------------(18)---------------------------
\begin{eqnarray}
R_s &=& \Big< \sin\theta_+ (r) \cos\phi_+ (r) \sin\theta_+ (0) \cos\phi_+ (0) \Big> \virg
\nonumber \\
R_c &=& \Big< \cos\theta_+ (r) \cos\phi_+ (r) \cos\theta_+ (0) \cos\phi_+ (0) \Big> \virg
\nonumber \\
R_{sc} &=& R_{cs} = \Big< \sin\theta_+ (r) \cos\phi_+ (r) \cos\theta_+ (0) \cos\phi_+ (0) \Big>
\point \nonumber \\
\label{eq:def-res}
\end{eqnarray}
%-----------------------------------------------------------

 In a  way similar to eqs.~(\ref{eq:RG}),
 eqs.~(\ref{eq:def-res}) are  calculated 
 where $r$ is replaced by  $\alpha \exp (l)$.
The corresponding  RG equations are given by 
\cite{Giamarchi-res,Cardy} 
(Appendix),
%---------------------------(19)-----------------------------------
\begin{subeqnarray}
 \frac{d}{dl}\:R_s(l) \hspace{-0.3cm} &=& \hspace{-0.3cm} - \, 
\Big[ \, (\, K_\rho(l) + y_{2\rho}(l))\theta(l_c - l) 
\nonumber \\
& & \hspace{0.5cm}                            + K_\sigma(l) + y_
\sigma(l) \, \Big] \, R_s(l) \,
\nonumber \\
& &                             + \theta(l_c - l) \, y_{1/2}(l) 
\, R_{sc}(l) \virg
 \\
& & \nonumber \\
 \frac{d}{dl}\:R_c(l) \hspace{-0.3cm} &=& \hspace{-0.3cm} -\, \Big[ 
\, (\, K_\rho(l) - y_{2\rho}(l))\theta(l_c - l) 
\nonumber \\
& & \hspace{0.5cm}                            + K_\sigma(l) + y_
\sigma(l) \, \Big] \, R_c(l) \,
\nonumber \\
& &                             + \theta(l_c - l) \, y_{1/2}(l) 
\, R_{sc}(l) \virg
 \\
& & \nonumber \\
 \frac{d}{dl}\:R_{sc}(l) \hspace{-0.3cm} &=& \hspace{-0.3cm} -\,
 \Big[ \, K_\rho(l) \, \theta(l_c - l) \,
                             + \, K_\sigma(l) + y_\sigma(l) \, \Big]
 \, R_{sc}(l) 
\nonumber \\
& &                             + \frac{1}{2} \, \theta(l_c - l)
 \, y_{1/2}(l) \, 
                                (\, R_s(l) + R_c(l) \, ) \virg
\nonumber \\
\label{eq:res-RG}
\end{subeqnarray}
%---------------------------------------------------------------
where $R_s(0) =R_c(0)=1/4$, $R_{sc}(0)=0$ and 
 $\theta (x) = 1 (0)$ for  $x>0 \, (x<0)$. 
The quantity $l_c$ denotes a scale at which 
 the coupling for the charge fluctuation develops well. 
  Then  the  region with  $l > l_c$
     is regarded as the strong coupling one where   
      the charge fluctuation  is  frozen. 
We take   $l_c = {\rm min}(l_{c1},l_{c2})$ 
 where the present numerical  calculation for the response function 
 gives a reasonable result with 
  a choice of $y_{1/2}(l_{c1}) = 1$ and $|y_{1/4}(l_{c2})| = 2$ 
   due to the first order RG.    
 Equation~(\ref{eq:res}) is estimated  from 
   eqs.~(\ref{eq:res-RG}),  which are calculated by using 
  the solution of eqs.~(\ref{eq:RG}).
 Although the quantity $F$  corresponds to 
    $R^{1/2}(|\vec{r}_1 - \vec{r}_2|=\infty)$, 
  we estimate   $F$ from the value of $R(l_m)$, i.e., 
%------------   (20)  -------------------------
\begin{eqnarray}
\label{eq:F_min}
  F  = R^{1/2}(l_m) 
        \virg   
\end{eqnarray}
%--------------------------
 where   $l_m$  denotes the  following characteristic length.  
 When $y_{\sigma}$ becomes of the order of unity due to the relevant 
   $y_{ps}$ and/or $y_{pc}$,  
 the spin gap is formed and 
    eq.~(\ref{eq:res}) as the function of $l$ 
    takes a minimum at $l=l_m$  
     due to the  first order RG.
Since $R(l)$ becomes invalid for $l \geq l_m$, we replace 
 $R(\infty)$ with $R(l_m)$.
Thus, $u$ is obtained  by  the following procedure.
 First, eq.~(\ref{eq:F_min}) is used for estimating $F$ 
 with various choices of $u$ 
 which gives the initial value of RG equations. 
 Next, $u$ is calculated  by relating  $F$ with $u$ through 
  eq.~(\ref{eq:SCE-p})  which is equivalent to 
 the Hellman-Feynman theorem.     
 Finally, $\zeta$ is determined to give the maximum of $u$.

Equation~(\ref{eq:phase Hamiltonian-RG}) shows that 
  the Peierls distortion is determined 
   by   $y_{ps}(l)$ and $y_{pc}(l)$
   with the optimum  $\zeta$.  
  Thus it is convenient to use a linear combination of 
    $y_{ps}$ and $y_{pc}$ terms. 
In the following section, we discuss the SP state by 
 rewriting $y_{ps}$ and $y_{pc}$ terms in 
eq.~(\ref{eq:phase Hamiltonian-RG}) 
as  
%------------------------------(21)----------------------------------
\begin{eqnarray}
   & &      \; \frac{v_F}{2 \pi \alpha^2} \, 
               \int dx \, 
     \Bigl[ - \, y_{p} \, \sin(\theta_+ + \zeta) \cos\phi_+ 
\nonumber \\
    & & \hspace{2cm}      - \, y_{pn} \, \cos(\theta_+ + \zeta) \cos\phi_+ 
     \Bigr]
\virg 
\label{eq:yp_ypn}
\end{eqnarray}
%-----------------------(22)----------------------------------------
\begin{eqnarray}
  \label{eq:yp}
y_p(l) \choose y_{pn}(l) &=&
  \pmatrix{
   \cos\zeta & \sin\zeta \cr
   -\sin\zeta & \cos\zeta \cr } 
 {y_{ps}(l) \choose y_{pc}(l)} \virg
\end{eqnarray}
%---------------------------------------------------------------
where $y_{p}(0)=4 \sqrt{2} \alpha u /v_F$ and $y_{pn}(0)=0$ .

%========================================================
\section{Dimerization vs. CO in the SP state} 
 First, we note the case where the distortion is absent 
  (i.e.,  $g=0$ and $x_d \not=0$). 
 The competition between  two kinds of umklapp scattering given by 
  $y_{1/2}$ term and $y_{1/4}$ term is expected with increasing $V$,
since the locking of $\left< \theta_+ \right> $
 of $y_{1/2}$ term and that of $y_{1/4}$ term in 
 eq.~(\ref{eq:phase-Hamiltonian1}) becomes different 
 for  $V > U/4$.
 There is a critical value, $ V = V_c$,  with a fixed $U$
     where $y_{1/2}$ becomes relevant (irrelevant) 
     for $V <V_{c} $ ($ V_{c}<V$)
      leading to the 1/2-filled state of dimerization
       (the  1/4-filled state of the CO).
%--------------  Fig. 1 --------------------------------
 The boundary between these two states 
  is shown by the dashed curve  in Fig.~1,
 \cite{Tsuchiizu,Ohta}
 where the dotted curve denotes a boundary for 
  $g=x_d=0$. 
 It turns out that   the region for the CO state is reduced 
 by the dimerization, i.e., $x_d$ 
 due to the competition between  $y_{1/4}$ and $y_{1/2}$ terms.
%============ fig1 ====================
\begin{figure}[tbp]
\begin{center}
 \vspace{2mm}
 \leavevmode
 \epsfysize=7.5cm\epsfbox{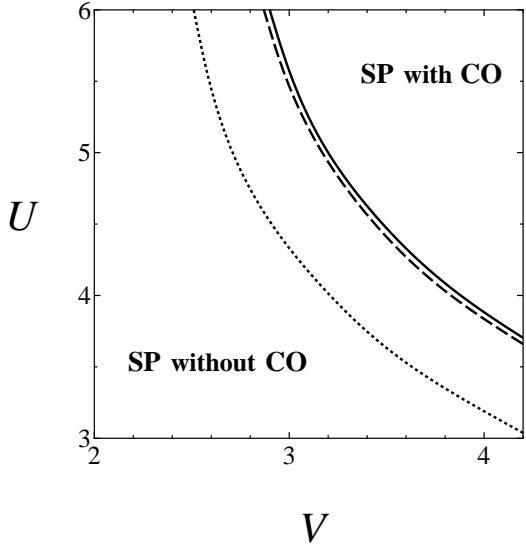}
 \vspace{-3mm}
\caption[]{
Phase diagram of an extended Peierls-Hubbard model 
 on the plane of $V$-$U$ with $x_d = 0.1$ and $g=0.1$ 
 where the solid curve is a boundary between 
 the spin Peierls (SP) state with CO (charge ordering) 
  and that without CO.  
A dashed curve (dotted) denotes the boundary for 
 $g \rightarrow 0$ and $x_d = 0.1$ ($g = x_d \rightarrow  0$). 
}
\end{center}
\end{figure}
%==========================================
 The case of  $V < V_c$ leads to  the Mott-Hubbard state 
  while the case of  $V_c < V$ exhibits   the CO state.
   The  ground state of the former state is  given by 
  $\left< \theta_+ \right> = \pi/4$  
   due to the relevant  $y_{1/2} (\rightarrow  + \infty)  $   
     while that of the latter state   is given by 
      $\left< \theta_+ \right> = \pi/2$ due to the relevant  
       $y_{1/4} (\rightarrow  - \infty)$.  
 For both states, 
   the charge excitation is gapfull and the spin excitation 
 is gapless. 

 Next, we consider the case  
  in the limit of small distortion where    
   the perturbational treatment of  the SP state is applicable  
     based on the state  with $g=0$.
 In this case,  the lattice distortion is determined   so as to obtain 
   the maximum value  of $ \sin (\theta_+ + \zeta)$ in
     eq.~(\ref{eq:yp_ypn}) since  $ y_{pn}(0)=0 $.   
 Thus  for $V < V_c$ and then $ \left< \theta_+ \right> = \pi/4$, 
  one finds  $\zeta = \pi/4$ leading to 2$\kf$ BOW and 4$\kf$ BOW
 (the state (a)), 
     while, for $V_c < V$ and then $\left< \theta_+ \right> = \pi/2$, 
       one finds  $\zeta = 0$ leading to  2$\kf$ CDW and 4$\kf$ CDW
 (the state (b)).
%---------------   Fig. 2 a,b,c ------------------------------
 These two states are shown in Figs.~2(a) and 2(b), respectively
  where the arrow denotes a spin and 
  the cross denotes the location for 
    the center of the singlet state.
%============ fig2 ====================
\begin{figure}[tbp]
\begin{center}
 \vspace{2mm}
 \leavevmode
 \epsfysize=7.5cm\epsfbox{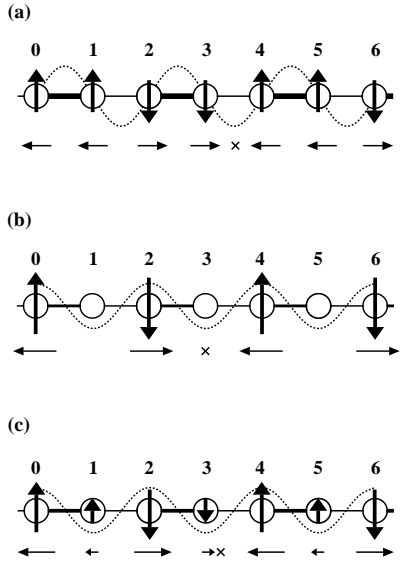}
 \vspace{-3mm}
\caption[]{
 Spatial variation of the SP state with  $\zeta=\pi/4$ for
 $V<V_c$ (the state (a)),  
 that with $\zeta=0$ for
 $V_c \ll V$ (the state (b)) and 
 that with $0 < \zeta < \pi/4$ 
for $V_c <  V$ (the state (c)), respectively.
 Dotted line denotes the $4 k_F$ charge density wave and  
  the vertical arrows (horizontal arrows) 
   denote spin ( the lattice displacement).
 The symbol, $\times$ ,is   
 the location for  the center of the singlet state.  
}
\end{center}
\end{figure}
%========================================== 
 However the presence  of the dimerization  
   suggests  a possibility of a  state given by Fig.~2(c)
 (the state (c)), 
  i.e., a mixed state with $0 < \zeta < \pi/4$. 
  In this section, we  examine not only the states of Figs.~2(a) and 2(b) 
  but also the  state of Fig.~2(c)
     by calculating  the optimum $\zeta$, which leads to    
        a  maximum of $u$ and  then a minimum of  
         the total   energy.

Now we calculate numerically the distortion, $u$, 
  by taking $t=1$ and $a=1$.
 We choose parameters as $U=5$, $x_d=0.1$ and $g=0.1$ 
  (i.e., $K \simeq 9.05$)  and 
   the results for other  parameters are   discussed later.
%------------------   Fig 3 a, b, c ------------------ 
 Figures~3(a) and 3(b), which are obtained from eqs.~(\ref{eq:RG}), 
  show   the  RG flow (i.e., $l$ dependence ) 
   for the case of small $V (< V_{c})$.
The magnitude of $u$
 is taken from the solution of eq.~(\ref{eq:SCE-p})
 where $u$ appears only in the initial value of RG 
 equations. 
%============ Figure 3 ====================
\begin{figure}[tbp]
\begin{center}
\vspace{2mm}
\leavevmode
\epsfysize=7.5cm\epsfbox{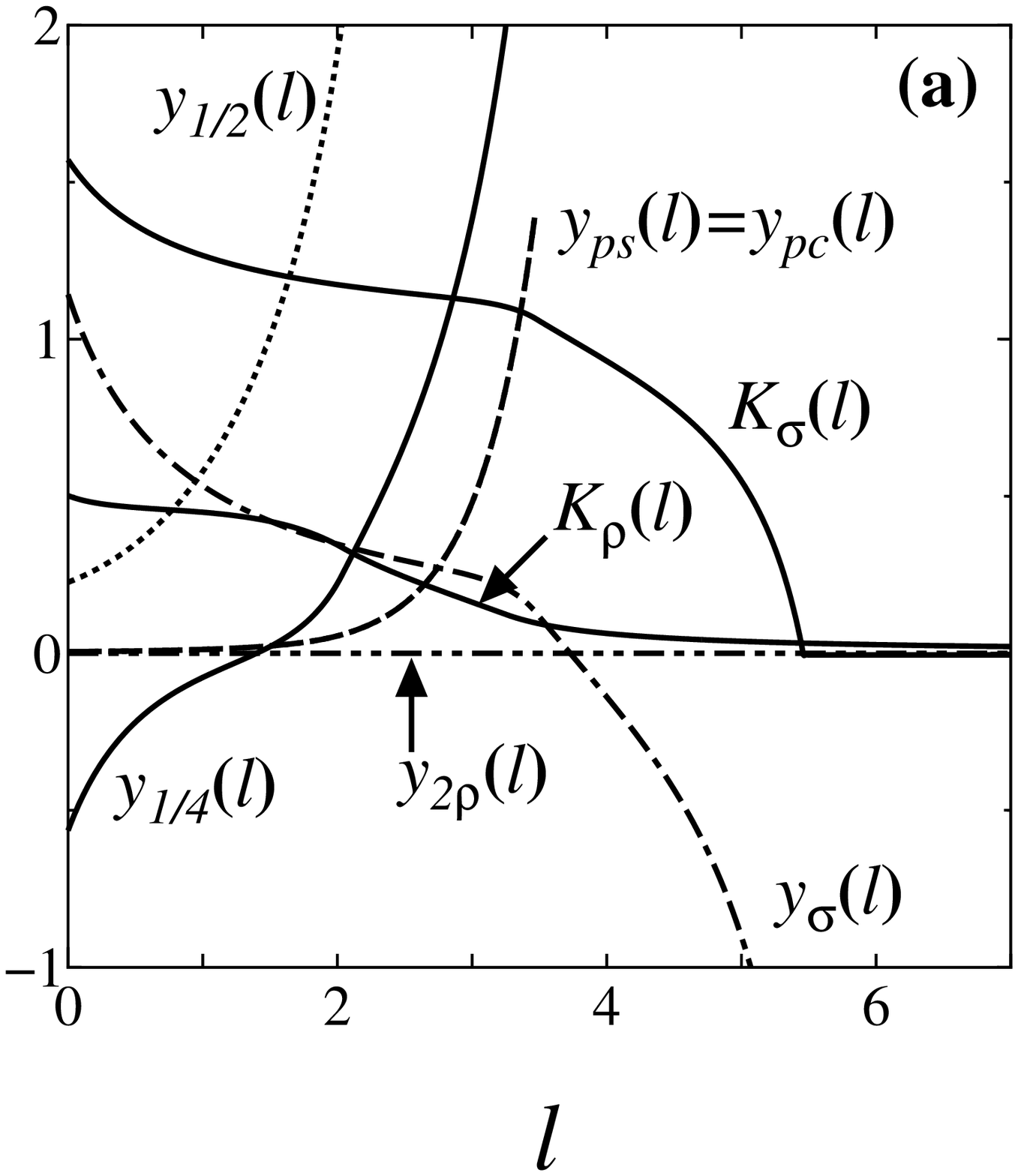}
\vspace{1cm}\\
\epsfysize=7.5cm\epsfbox{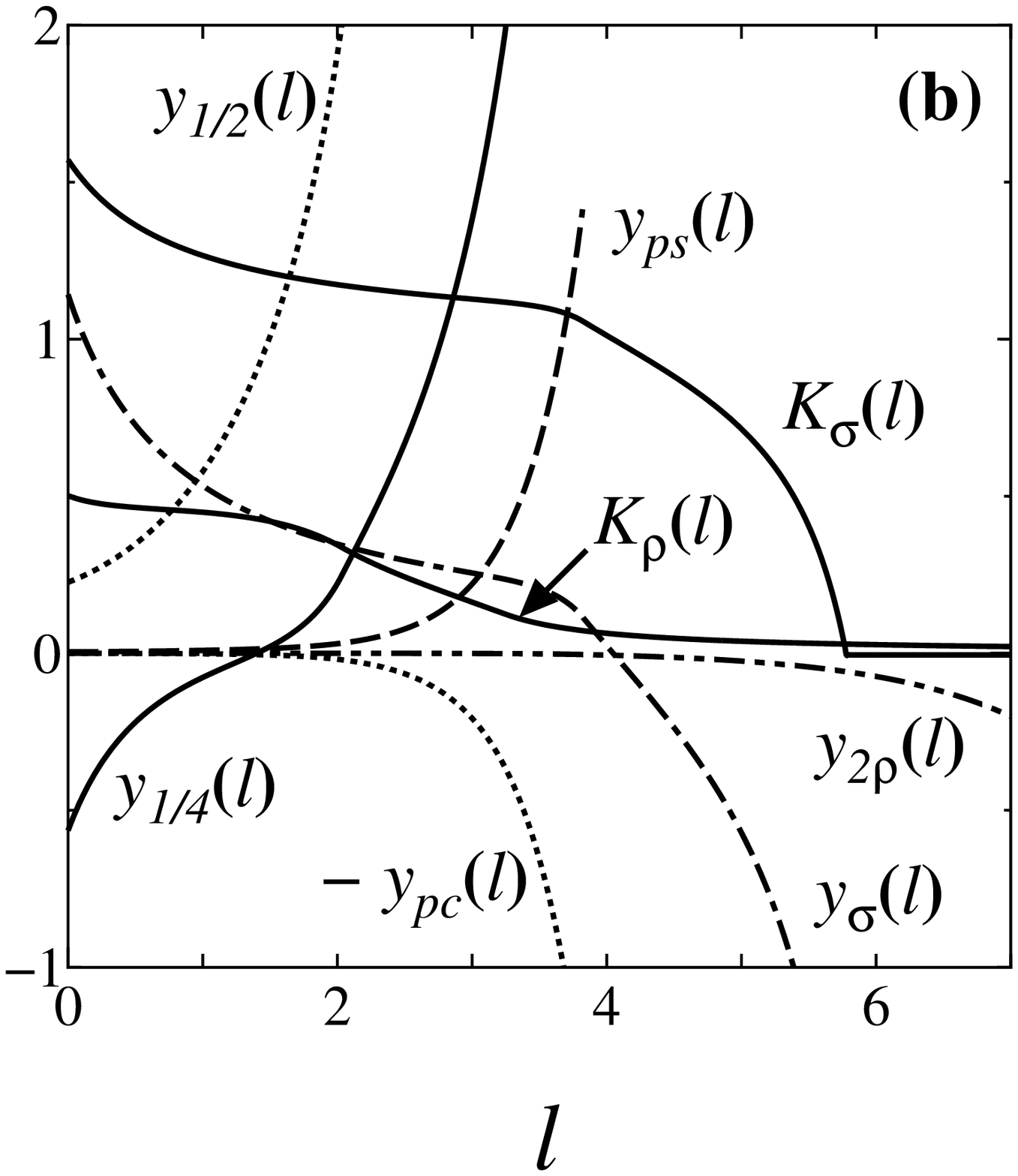}
 \vspace{-3mm}
\caption[]{
The $l$ dependence of 
 $K_\rho(l),K_\sigma(l),y_{1/4}(l),
   y_{1/2}(l), y_{2\rho}(l), \\
    y_\sigma(l),y_{ps}(l)$ and $y_{pc}(l)$
           for $V=2$ with 
 $U=5, x_d=0.1$ and $g = 0.1$  
  where   $\zeta=\pi/4$ $(u=0.0029)$ (a) 
       and $\zeta=0$ $(u=0.0018)$ (b). 
 Both cases  lead to the bond order.
 }
\end{center}
\end{figure} 
%============================================
 The most relevant quantity is $y_{1/2}$ 
   and then $y_{1/4}(l)$ changes the sign from a negative value 
   to a positive value with increasing $l$. 
 Figure~3(a) 
  shows  $\zeta=\pi/4$ 
 and      $\left< \theta_+ \right> =\pi/4$ 
  due to    $y_{ps}(l) = y_{pc}(l)$  
 while Fig.~3(b) shows 
 $\zeta= 0$ and   $\left< \theta_+ \right> = \pi/4$. 
Since  $\left< \theta_+ \right> = \pi/4$  
  regardless the initial condition of $\zeta$, 
  both Figs.~3(a) and 3(b) show 4$\kf$ BOW  and 2$\kf$ BOW. 
By substituting   $y_{ps}(l)$ and $y_{pc}(l)$ of Figs.~3(a) and 3(b)
 into eq.~(\ref{eq:yp}), 
  it turns out that Fig.~3(a) shows the largest $y_p(l)$ indicating 
  the ground  state with $\zeta=\pi/4$.
  The relevant $y_{1/2}$ results in 
   the charge gap  determined by the dimerization, which  
    leads to an effectively half-filled band. 
 A spin gap  is  obtained  
      from the relevant $y_{\sigma}$, which  comes from   
        $y_{ps}$ and $y_{pc}$ as seen from eq.~(\ref{eq:RGf}).  
 The spin gap is much smaller than the charge gap
     since   $y_{1/2}(l)$ increases rapidly compared with $y_{p}(l)$. 

%-------------------   Fig 4 a, b, c -------------------------------
 The  RG flows    for the case of large $ V (>V_{c}) $ are  shown 
  in Figs.~4(a), 4(b) and 4(c) 
 where $u$ is chosen from the solution of 
 eq.~(\ref{eq:SCE-p}).  
%============ Figure 4 ====================
\begin{figure}[tbp]
\begin{center}
 \vspace{2mm}
 \leavevmode
\epsfysize=7.5cm\epsfbox{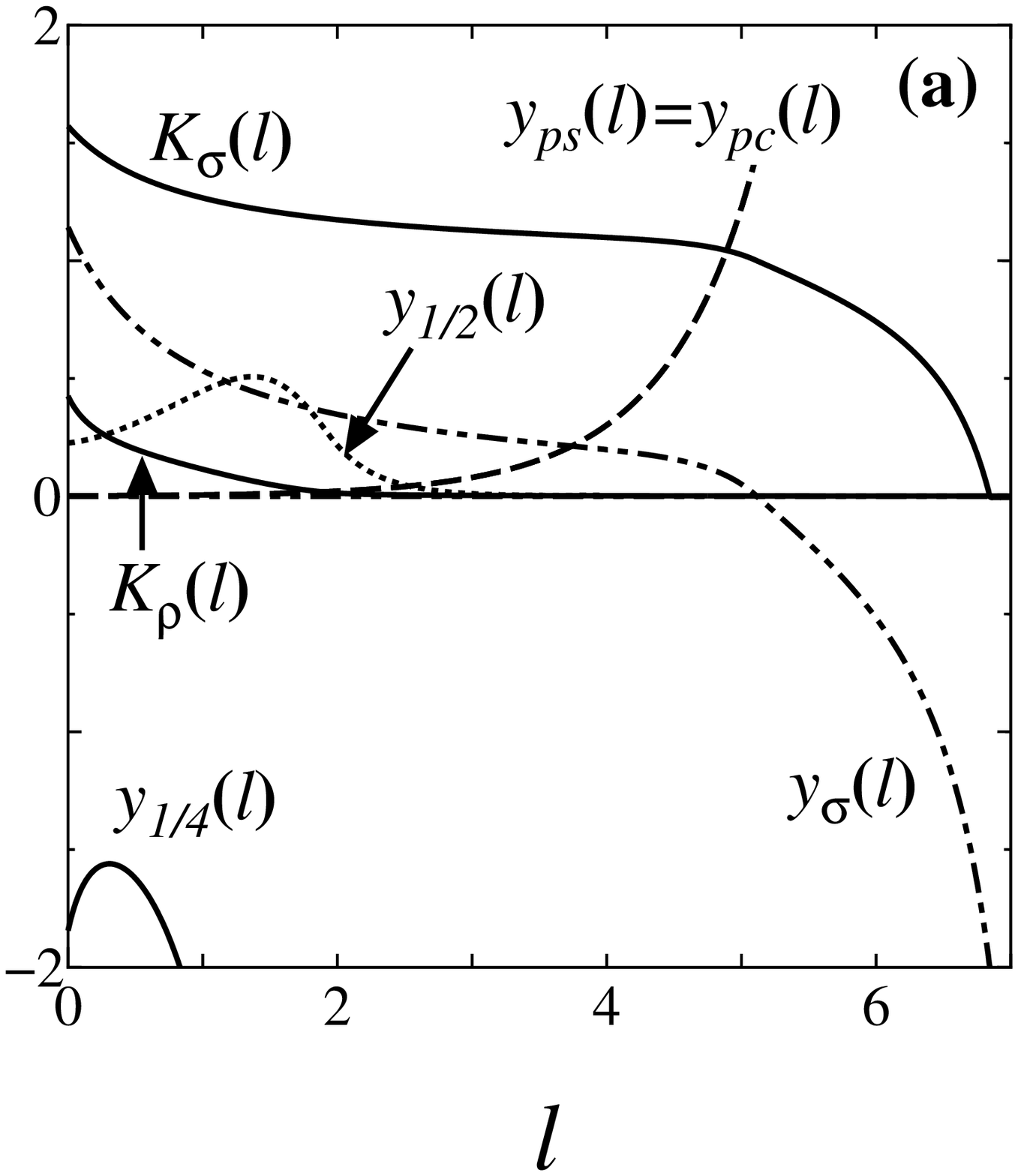}
\vspace{0.5cm}\\
\epsfysize=7.5cm\epsfbox{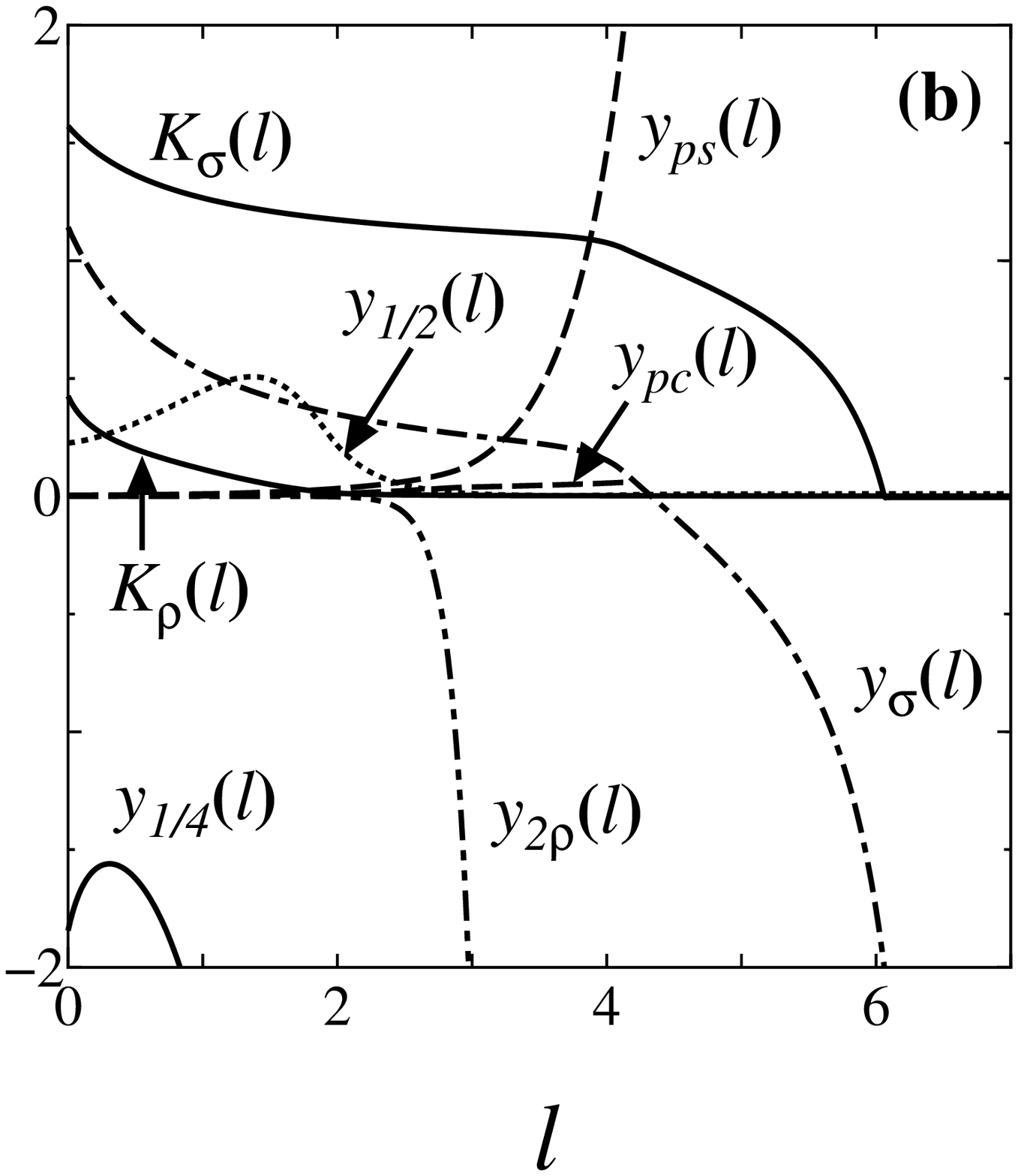}
\vspace{0.5cm}
\epsfysize=7.5cm\epsfbox{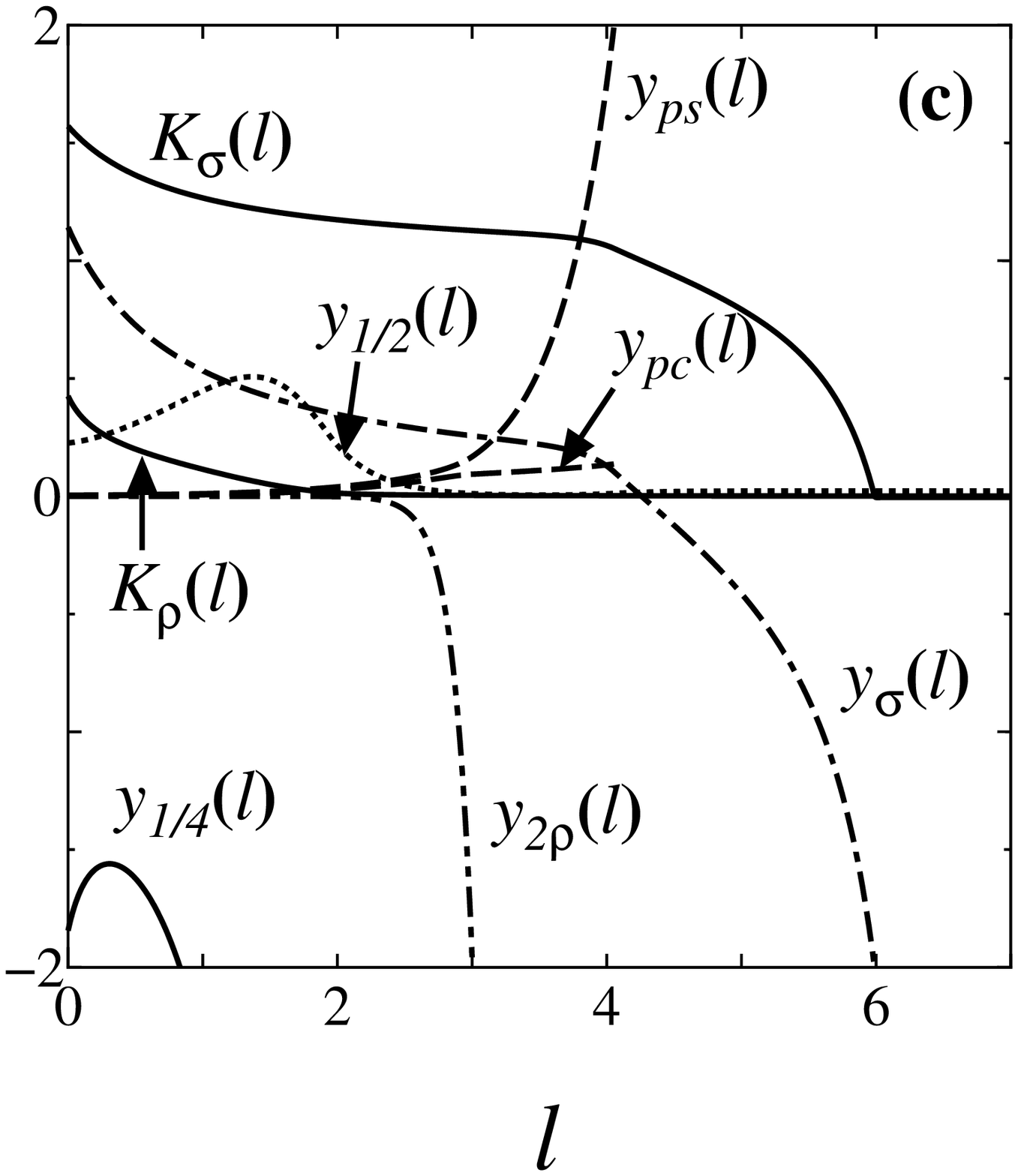}
 \vspace{-3mm}
\caption[]{
The $l$ dependence of 
 $K_\rho(l), K_\sigma(l), y_{1/4}(l),
   y_{1/2}(l), y_{2\rho}(l),$
 $ y_\sigma(l),y_{ps}(l)$ and $y_{pc}(l)$
  for  $V=3.7$ with 
  $U=5, x_d=0.1$  and $g = 0.1$ 
  where  
  $\zeta=\pi/4$ $(u=0.0012)$(a),
   $\zeta=0$ $(u=0.0017)$ (b) and 
    $\zeta=0.6 \times \pi/4$ $(u= 0.0020)$ (c),
  respectively.
 In  the first panel (a), $y_{2\rho}(l) = 0$. 
 All these cases lead to the CO due to relevant $y_{1/4}(\to - \infty)$.  
 }
\end{center}
\end{figure} 
%========================================== 
 In all these figures, 
   $y_{1/2}(l)$ becomes irrelevant 
  due to  the relevant $y_{1/4}(l) ( < 0)$ 
   indicating   4$\kf$ CDW  with  $\left< \theta_+ \right> =\pi/2$.
 Figure~4(a)  with   $\zeta =\pi/4$  
   shows   
      $y_{ps}(l) = y_{pc}(l)$  which implies 
 $y_{p}(l) > 0$  and $y_{pn}(l) = 0$.
 Figure~4(b) with $ \zeta = 0$ 
  also  shows  
   the  relevant $y_{p}$ and  the irrelevant  $y_{pn}$.
 From the comparison of $y_{p}$ of Fig.~4(a) and that of Fig.~4(b), 
 it is found that 
      Fig.~4(b)  gives  the distortion  larger than  Fig.~4(a).  
 However the larger distortion is expected from 
      the   state  shown by Fig.~2(c) 
       with  $ 0 < \zeta <  \pi/4 $, 
        since  the dimerization breaks the symmetry around  the 
       distortion  given by the cross of  Fig.~2(b).
 Actually  
 an example of such a state is  shown in Fig.~4(c) 
 where $y_{p}(l)$ increases rapidly compared with that of 
    Fig.~4(b).
 For   Fig.~4(c),   
 $y_{pn}(l)$ of  eq.~(\ref{eq:yp})
 decreases from zero to a negative value,  
 while 
    $y_{pn}(l)\simeq  0$  for  Figs.~3(a) and 4(b).
Thus  the state for   $V > V_c$ is given by 
  Fig.~4(c) with   $\zeta > 0$ 
  due to  the energy gain from both $y_{p}$ term and $y_{pn}$ term. 

The distortion $u$ is determined by   eq.~(\ref{eq:SCE-p}), 
   with   $F$ estimated  from  
  the response function $R(l)$ 
  where  eq.~(\ref{eq:res})   is  calculated from 
         eqs.~(\ref{eq:def-res}) and (\ref{eq:res-RG}).
%----------------------   Fig 5 -------------------------------------
 In Fig.~5, 
  the response function $R(l)$ for $\zeta=\pi/4$ 
  is shown  for 
   the case of small $V (< V_{c})$ (solid curve) 
     and large $V (>V_{c})$  (dotted  curve), respectively, 
   where  $V_{c} = 3.2$.
 %============ Figure 5 ====================
\begin{figure}[tbp]
\begin{center}
 \vspace{2mm}
 \leavevmode
 \epsfysize=7.5cm\epsfbox{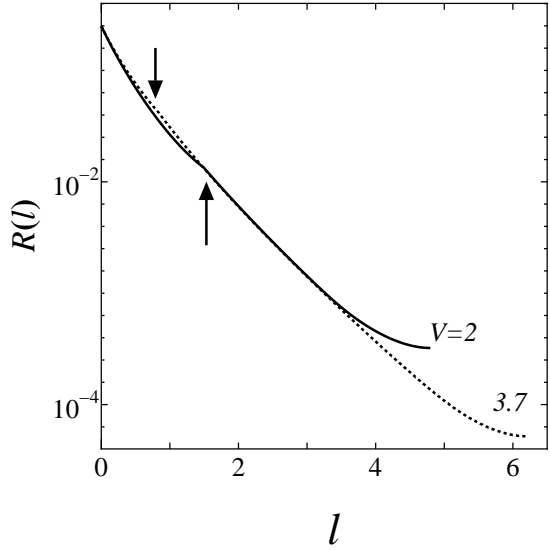}
 \vspace{-3mm}
\caption[]{
The $l$ dependence of $R(l)$  for  $\zeta=\pi/4$
 with  V=2 (solid curve), 3.7 (dotted curve)
  where  $U=5 , x_d = 0.1$ and  $g = 0.1$. 
 The arrow denotes $l (= l_c)$, at which   
 the charge fluctuation is frozen.
The end point of $R(l)$ denotes  
 a minimum of $R(l)$  at   $l =l_m$.
}
\end{center}
\end{figure}
%==========================================
 When  the response function enters into  the region 
   of forming the charge  gap,
   we stop the RG treatment for  the charge  
    fluctuation and take into account  only the spin fluctuation 
  by keeping 
   the   relevant coupling of  eq.~(\ref{eq:phase Hamiltonian-RG})
      as a constant for  $l>l_c$. 
  The arrow indicates a location for  $l=l_c$ 
    at which the charge fluctuation becomes frozen
          due to  a formation of the charge gap. 
 Finally, we stop  calculating  
  the RG equations of eqs.~(\ref{eq:res-RG})
  when it takes a minimum at $l=l_m$ (end point) 
 corresponding to   the SP state. 
It is expected that  the difference between $R(l_m)$  
 and the correct  value at the  long distance is small  
since
 the formation of the spin gap occurs at $l \simeq l_m$  
  due to the SP state.   
%-------------------  Fig 6  ---------------------------
Figure~6 shows $u$ dependence of 
  $F (= R^{1/2}(l_m))$, which is obtained from Fig.~5
   with fixed $\zeta = \pi/4$ and  some choices of $V$.
%============ Figure 6 ====================
\begin{figure}[tbp]
\begin{center}
 \vspace{2mm}
 \leavevmode
 \epsfysize=7.5cm\epsfbox{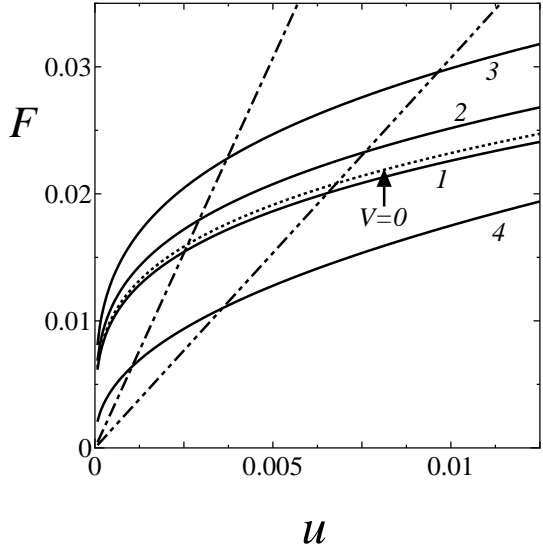}
 \vspace{-3mm}
\caption[]{
 The quantity $F$  of eq.~(\ref{eq:SCE-p}), 
 as the function of $u$ for $\zeta =\pi/4$
 with  $U=5 , x_d = 0.1$ and $V = 1, 2, 3 $ and $4$. 
 The dash-dotted ( dash-two dotted) line denotes 
  $(\sqrt{2} a/g v_F) u $ of eq.~(\ref{eq:SCE-p}) 
   for   $g =$ 0.1 (0.2), where 
  the solution of eq.~(\ref{eq:SCE-p}) is obtained from
     the intersection.    
}
\end{center}
\end{figure}
%==========================================
 The quantity $F$ increases monotonically  with increasing 
  $u$ where the power law as a function of $u$ is expected 
   in the presence of the interaction as seen also for 
   the  half-filled case.
\cite{Sugiura1}  
 It turns out that  $F$ takes a maximum as a function of $V$. 
 Such a maximum  originates in a fact that 
  the  SP state  with $\left< \theta_+ \right>=\pi/4 $ 
  is replaced by the CO with   
      $\left< \theta_+ \right>=\pi/2 $ in the case of  $V > V_c$. 

%--------------  Fig 7 ---------------------------------
 Figure~7 is the  main result of the present calculation where 
   the solution $u$ for eq.~(\ref{eq:SCE-p}) is obtained from 
 the intersection in Fig.~6.
%============ Figure 7 ====================
\begin{figure}[tbp]
\begin{center}
 \vspace{2mm}
 \leavevmode
 \epsfysize=7.5cm\epsfbox{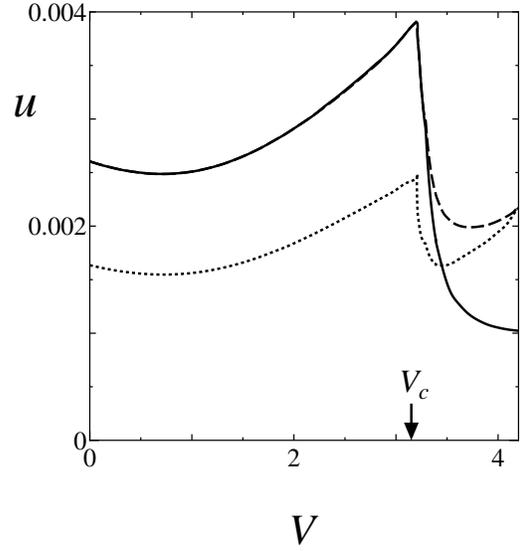}
 \vspace{-3mm}
\caption[]{
The $V$ dependence of $u$ 
  for  $\zeta=\pi/4$ (solid curve), 
   $\zeta=0$ (dotted curve) 
    and the optimum value of $\zeta$ (dashed curve)
     where  $U=5 , x_d=0.1$ and $g=0.1$.
}
\end{center}
\end{figure}
%========================================== 
  The  solid curve   corresponds to $u$ 
  for $\zeta=\pi/4$ (Fig.~2(a)) 
  while the dotted curve denotes   $u$ for 
     $\zeta=0$ ( Fig.~2(b)).    
   The latter case of  $\zeta=0$ 
  is also calculated  in a way similar to Fig.~6.  
Both curves exhibit  a  maximum at $V=V_c$ shown by the arrow
   where the CO state appears for $V > V_c$.
The maximum at $V=V_c$ originates in the competition between 
 the dimerization and the CO, which is comprehended as follows.
 The numerical results together with 
 eqs.~(\ref{eq:RGc}) and (\ref{eq:RGd})
  show that   
    $y_{1/2}$ becomes irrelevant 
     and $y_{1/4}$ becomes relevant for    $V_c < V$.
 Since the length  $l$ for the  minimum of $R(l)$ corresponding to  $u$ 
  is nearly equal to $l$ with $|y_{\sigma}(l)|=2$ ,
   the distortion is essentially determined by  
    $[y_{ps}^2(l) + y_{pc}^2(l)]^{1/2}$ 
    in  eq.~(\ref{eq:RGf}).
 Equations~(\ref{eq:RGg}) and (\ref{eq:RGh}) 
 show  that 
  $y_{ps}(l)$ and $y_{pc}(l)$ are 
   determined by the umklapp scatterings, 
    $y_{1/2}(l)$ and $y_{2\rho}(l)$,  where  
    $y_{2\rho}$    is induced by 
    $y_{1/4}$ for $V_c < V$ (eq.~(\ref{eq:RGe})).
 Thus  the sudden decrease of $u$ for $V>V_c$     
  originates in the fact that  
  the increase of $y_{1/2}(l)$ for $V < V_c$ is much larger than 
  that of $y_{2\rho}$ for $V_c<V$,
  close to $V = V_c$.
 The  increase of $u$ with $V$ just below $V_c$ comes from 
 the decrease of $K_{\rho}$,
  which  results  in the suppression of the 
   charge fluctuation.   
From the comparison of the solid curve with the  dotted curve,
   the maximum of $u$ for $V < V_c$ is given by  
    the SP state with $\zeta = \pi/4$ (solid curve) 
 shown in   Fig.~2(a),
\cite{Sugiura2}
    while $u$ with $\zeta=0$ (  dotted curve) becomes larger than 
    that of the solid curve 
      for large $V$  $( >  V_c) $. 
 The first order transition might be   expected 
  between   these two states 
   if   the dimerization were absent.
\cite{Seo_Meeting}  
 However,
  the   mixed state  shown in Fig.~2(c) is rather promising 
  for $V > V_c$ 
   since  the dimerization does exist
       even  for the SP state with the CO.
 The dashed curve denotes  $u$ of such a mixed state, 
  which is obtained by 
  choosing the optimum $\zeta$  leading to the maximum of $u$.
  The difference between the dashed curve and the dotted curve 
   becomes small for $V \gsim 4$.
 With increasing $V$ for $V > V_c$, the distortion $u$ increases again   
 since the state of Figs.~4(b) and (c) are compatible with 
 the CO, and   
  $K_{\rho}$ keeps decreasing with increasing  $V$.
 The minimum of $u$ for $V < V_c$ comes from 
 the change of the sign of $y_{1/4}(0)$ as the function of $V$.  
Thus we obtain the  transition from 
  the state of Fig.~2(a) to  that of Fig.~2(c)  at  $V = V_c$.

%-----------------   Fig 8 -------------------------------------
  In Fig.~8, the $V$ dependence of $\zeta$ for the mixed state 
  (Fig.~2(c)) is shown 
   where  $x_d$ = 0.1(solid curve) and $x_d =$ 0.2(dotted curve).
%============ Figure 8 ====================
\begin{figure}[tbp]
\begin{center}
 \vspace{2mm}
 \leavevmode
 \epsfysize=7.5cm\epsfbox{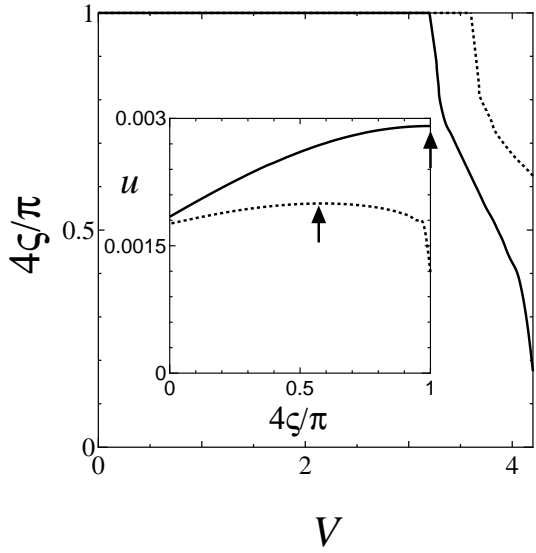}
 \vspace{-3mm}
\caption[]{
The $V$ dependence of the optimum  $\zeta$
 for $x_d=0.1$ (solid curve) 
     and $x_d=0.2$ (dotted curve) 
   where  $U=5$ and  $g = 0.1$. 
 The inset denotes the corresponding $\zeta$ dependence of $u$ 
    for $V=2$(solid curve) and $V=3.7$(dotted curve) for $x_d=0.1$ where  
    the arrow denotes the location for the optimum $\zeta$.
}
\end{center}
\end{figure}
%==========================================
 The quantity $\zeta$ begins to decrease  from $\pi/4$ 
   at a critical value and reduces  to zero monotonically.
 Within the numerical accuracy,
     the onset of the mixed state
     is given by $ V =V_c$  
  indicating the second-order  transition.
  With increasing $x_d$,
  $V_c$  increases  
   since the SP state with $\zeta=\pi/4$ is enhanced by 
   the  dimerization ($\propto y_{1/2}$). 
 The inset shows $u$ as a function of $\zeta$ 
  where the location of  the optimum $\zeta$ is shown by the arrow.
   For $V < V_c$, $u$  as a function of   $\zeta$
   increases  monotonically   while $u$  for $V_c < V$
   takes a maximum in the interval range of    $0<\zeta<\pi/4$.   

%------------------   Fig 9  --------------------------------
 The energy, $\vf \alpha^{-1} \exp (-l_m)$, corresponding to 
   the SP state is nearly equal to that of  
   the freezing   of the spin fluctuation 
 while   the charge fluctuation is frozen at   higher energy 
      as seen from Fig.~5 (the location indicated by the arrow). 
%============ Figure 9 ====================
\begin{figure}[tbp]
\begin{center}
 \vspace{2mm}
 \leavevmode
 \epsfysize=7.5cm\epsfbox{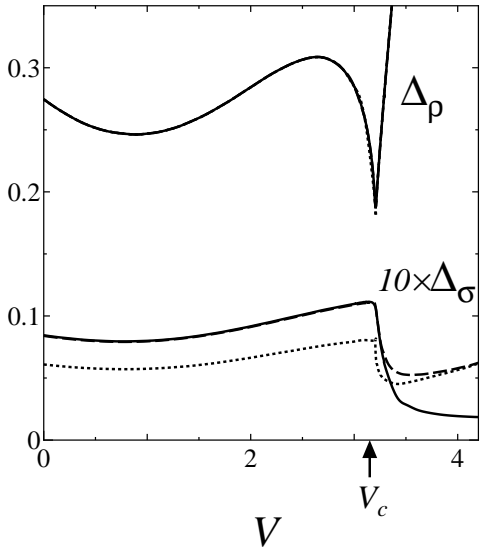}
 \vspace{-3mm}
\caption[]{
The $V$ dependence of the charge gap  $\Delta_\rho$ 
 and  the spin gap $\Delta_\sigma$
 for  $\zeta=\pi/4$ (solid curve), 
   $\zeta=0$ (dotted curve) 
   and the optimum $\zeta$ (dashed curve) 
 where   $U=5,x_d=0.1$ and  $g = 0.1$.
 The difference among three curves of $\Delta_\rho$
  is   invisible.   
 The spin gap, which  is multiplied  by 10, 
  is similar to $u$ in Fig.~7. 
}
\end{center}
\end{figure}
%==========================================
 Figure~9 shows   the  charge gap,  $\Delta_{\rho}$,
       and the spin gap,  $\Delta_{\sigma}$, which correspond to  
    Fig.~7.
These gaps are   calculated from  
 $\Delta_{\rho} = \omega_c \exp [-l_{\rho}]$ and  
 $\Delta_{\sigma} = \omega_c \exp [-l_{\sigma}]$
 where 
   $l_{\rho} = {\rm min} (l_{\rho 1}, l_{\rho 2})$ 
   with $y_{1/2}(l_{\rho 1}) = 2, |y_{1/4}(l_{\rho 2})| = 3$
    and $|y_{\sigma}(l_{\sigma})| = 2$.
 The cutoff is chosen as  
  $ \omega_c = 2.2$, by noting that 
  $\omega_c =$ 5.3 for the half-filled case \cite{Lieb}
   and the ratio of the bandwith of half-filling  and 
    that of quarter-filling is 0.41.
\cite{Tsuchiizu-1/2}
 The length $l_{\rho}$ is essentially the same but 
  is slightly larger than $l_{c}$ in Fig.~5,
    which is chosen for the convenience  of 
     the numerical calculation of  $R(l)$. 
   The charge gap   is much larger than the spin gap,  
      which  is nearly equal to  $u$.   
 The dip of the charge gap and the cusp of the spin gap, 
  which are found at  $V = V_c$, originates in the fact that  
    the fixed point of $y_{1/4}(l)$ changes from 
     $+ \infty$ to $- \infty$       at $V =V_c$ 
    (or equivalently              
             from $+\infty$ to 0   for $y_{1/2}(l)$).

%------------------- Comment ----------------------------
Finally, we discuss  the relation between the  method of 
 obtaining the optimum $\zeta$ and that of 
   $\langle \partial H / \partial \zeta \rangle = 0$ shown  just above 
 eq.~(\ref{eq:SCE-p}). 
 The estimation of the latter one for the state of Fig.~2(c)
  is very complicated 
 within the present scheme of RG method. 
  However,  the  condition given by
   $\langle \partial H / \partial \zeta \rangle = 0 $  
 (i.e.,  $\langle \cos(\theta_+ + \zeta) \cos \phi_+ \rangle = 0$)
   seems to be compatible  with the former result with 
  $0 < \zeta < \pi/4$ 
 if we note the following property of 
 the quantum fluctuation  of $\theta_+$ for  $ V >V_c$. 
 The fluctuation with low energy (i.e., large $l$)
 exists  around 
  $\langle \theta_{+} \rangle = \pi/2$  due to the fixed point 
 of the CO state
 while 
 the fluctuation with high energy (i.e., small $l$) 
 exists  around  
 $\langle \theta_{+} \rangle = \pi/4$  due to 
 the dimerization  as seen from $y_{1/2}(l) \not=0$ 
 for small $l$ in Fig.~4(c).  
Further, it should be noted that  
 such a competiton of the dimerization and the CO becomes noticeable
  for  the state with 
 $V$ just above $V_c$. 
 
%----------------------------------------------

\section{Discussion}
  Using an extended Peierls-Hubbard model where 
   the presence of the dimerization is assumed 
    as a model of  organic conductors,
  we have examined the SP state and obtained 
   the following results.
With increasing nearest-neighbor interaction, $V$,
 the transition from 
 the SP state of Fig.~2(a)   into that of Fig.~2(c) 
 occurs   at $V = V_c$ corresponding to the appearance of the CO 
 where  the  transition is of second-order due to the dimerization.
 The competition 
  of  the dimerization with  the CO  results in 
  the maximum of $u$ at $V=V_c$ 
  and  the rapid decrease of $u$ for $V$ just above $V_c$. 

We have also examined the SP state with other  parameters.
The calculation of $u$ similar to Fig.~7  
  for $U=$4,5 and 6 shows that  
  $u$ as the function of $U$ decreases monotonically in the  region 
   around the maximum (i.e., $V=V_c$) 
 as found  for the half-filled case.
\cite{Sugiura1} 
 The dimerization $x_d$  increases   
   $V_c$  as seen from   Fig.~1 where  
    $V_c$ for $x_d=0.1$ (solid curve) 
    is much larger than that for $x_d \rightarrow 0$ (dotted curve).
\cite{Mila,Yoshioka_JPSJ00}
 The SP  state  without the CO is also enhanced in the presence of 
   $g$ since the SP state of Fig.~2(a) is compatible with 
   the Mott-Hubbard state.   
 The magnitude $u$ of Fig.~7 obtained by  $g=0.1$  is reasonable for  
  the organic conductor in which  $t \sim 2000 K$
\cite{Ducasse}
 and $u (\sim T_{\rm SP}) \sim$ 10 K leading to 
$u/t \sim$ 0.005.
\cite{Bechgaard}  
The small effect of $g$ on  $V_c$ is found 
   from the comparison of  
    the solid curve with the dashed curve in Fig.~1
    and from the fact that 
    the enhancement of  $V_c$  is about 0.1 even for $g=0.5$.  
However, $u$ is strongly enhanced by $g$. 
%--------------------   Fig 10 -------------------------------------
%============ Figure 10 ====================
\begin{figure}[tbp]
\begin{center}
 \vspace{2mm}
 \leavevmode
 \epsfysize=7.5cm\epsfbox{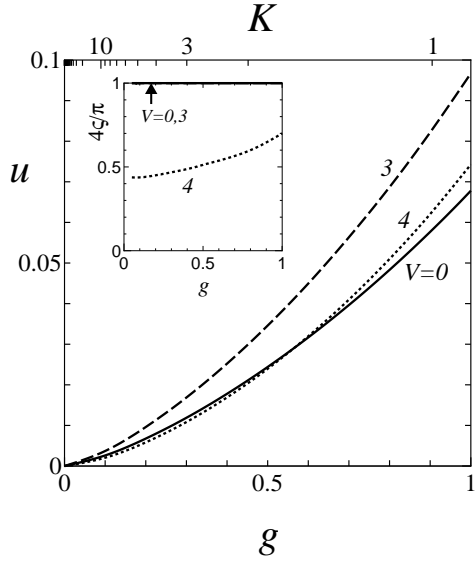}
 \vspace{-3mm}
\caption[]{
The $g$ dependence of $u$ for 
 $V =$ 0, 3 and 4 with   $U=5$ and $x_d=0.1$.
The inset denotes the optimum $\zeta$ corresponding to 
 the main figure.
}
\end{center}
\end{figure}
%=================================================
The  $g$ dependence of $u$  is shown   
  in Fig.~10  with some choices of $V$.
 With increasing $g$,  
  $u$  increases with a power law due to the presence of $U$     
  while $u \propto \exp (-1/g)$  for the conventional Peierls state 
 (i.e., $U=V=0$).
The $V$ dependence of $u$ is weaker than the $g$ dependence of $u$.
Thus the critical value of $V_c$ in Fig.~7 does not depend much on $g$
 indicating also  the small effect  of  $g$ on  the boundary in  Fig.~1. 

We comment on our treatment of the first order RG. 
 We obtained the cusp of $u$ in Fig.~7 and 
  the  dip of $\Delta_{\rho}$ in Fig.~9. 
 Such an anomaly  originates in the fact that  
   the $y_{1/2}$ term competes with  the $y_{1/4}$ term. 
This competition  has a common feature with  the  Ising transition,
\cite{Fabrizio,Tsuchiizu_Orignac} 
which leads to the vanishing of $\Delta_{\rho}$ at $V=V_c$.
 Thus it is expected that $u$ in Fig.~7 remains finite but 
 the tangent   becomes infinity.

Finally,   based on  the   result of Fig.~7 and Fig.~9, 
  we comment on the SP state of organic conductors
  with the dimerization
  by noting that  $u \propto T_{\rm SP}$, and   pressure 
    decreases   $V/t$ due to  the increase of  $t$. 
 The experiment on  (TMTTF)$_2$AsF$_6$
\cite{Zamborszky} 
   is interpreted  as follows.  
The SP state under low pressure  ( $<$ 0.15 GPa), which 
   shows   the increase of $T_{\rm SP}$ and the decrease 
of $T_{\rm CO}$  with increasing pressure, 
     corresponds  to 
     the mixed state of Fig.~2(c) obtained  for $V$ just above  $V_c$. 
 The SP state  under high pressures ( $>$ 0.15 GPa ) 
  showing  the decrease of 
  $T_{\rm SP}$  as the function of  pressure, corresponds to 
  the  SP state of  Fig.~2(a). 
 Further we note 
  the SP state of (TMTTF)$_2$X  with X = PF$_6$ and AsF$_6$ 
   have  $T_{\rm CO} \simeq $ 62K and 103K and 
   $T_{\rm SP} \simeq$  18K and 11K, respectively at ambient pressure.
\cite{Zamborszky}  
 These two SP states  may be described by  the mixed state of Fig.~2(c)    
    since 
  $V/t$ of (TMTTF)$_2$AsF$_6$ is   larger than  
  (TMTTF)$_2$PF$_6$ due to the effective pressure.

%-----------------------------------------------
\section*{ Acknowledgments }
The authors are grateful to M. Ogata and  H. Seo 
for useful  comments  on  the state  of Fig.~2(b).
They  also thank  S. Brasovskii and A. Furusaki for valuable  
 discussions. 

\appendix

\section{Renormalization Group Equations}

First, we derive the renormalization group (RG) equations 
 for $H^{\rm{eff}} = H_0 + H_1^{\rm{eff}}$ 
  (eq.(\ref{eq:phase Hamiltonian-RG})), 
   using the response function \cite{Giamarchi} 
    given by
%===================(A1)=========================================
\begin{eqnarray}
\hspace{-0.5cm} R(\vec{r}_1 - \vec{r}_2)
&=& \left \langle \, T_\tau e^{{\rm i} \theta_+(\vec{r}_1)}
                            e^{- {\rm i} \theta_+(\vec{r}_2)}\, 
                            \right \rangle_{H^{\rm{eff}}} \nonumber \\
&=& \frac{1}{\langle S_I \rangle_0} \left \langle \, T_\tau 
    e^{{\rm i} \theta_+(\vec{r}_1)} e^{- {\rm i} \theta_+(\vec{r}_2)} 
    S_I \, \right \rangle_0 
\label{eq:Response} \point
\end{eqnarray}
%=====================================================================
$T_\tau$ is time ordering operator, $\vec{r} = (x,v_F \tau)$ 
 and $\langle \;\;\; \rangle_0$ denotes 
  the average over $H_0$ .
In the absence of $H_1^{\rm{eff}}$, 
  $R(\vec{r}_1 - \vec{r}_2) = \exp\{-K_\rho U(\vec{r}_1 - \vec{r}_2)\}$ 
   with $ U(\vec{r}) = \ln(|\vec{r}|/\alpha) $ .
$S_I = T_\tau \exp[- \int d^2 {\tilde r} {\tilde H}_1^{{\rm eff}}]$ 
 with $H_1^{{\rm eff}} = (v_F/2 \pi \alpha^2) \int dx 
  {\tilde H}_1^{{\rm eff}}$ and 
   $d^2{\tilde r} = dx \, v_F d \tau / 2 \pi \alpha^2$.

We expand the nonlinear terms of $H_1^{\rm{eff}}$ treating 
 as perturbation and rewrite the response function as 
    $R(\vec{r}_1 - \vec{r}_2) = 
     \exp\{-K^{\rm{eff}}_\rho U(\vec{r}_1 - \vec{r}_2)\}$.
$K^{\rm{eff}}_\rho$ is written by 
  $y_{1/4},y_{1/2}, \cdot \cdot \cdot$. 
   Assuming the scale invariance of response function, 
    one obtain the RG equations.

By expanding ${\tilde H}_1^{{\rm eff}}$ up to third order, 
 the response function is written as
%======================(A2)======================================
\begin{eqnarray}
& & R(\vec{r}_1 - \vec{r}_2) 
= \exp\left( \, - f_\rho(1,2) \, \right) \, \Bigg[ \, 1 + \sum_{\epsilon=\pm} 
    \int d^2{\tilde r}_3 d^2{\tilde r}_4 \nonumber \\
& &
\times \, \Big\{ \, \frac{1}{8} y_{1/4}^2 \exp\left( -16 f_\rho(3,4) \right) 
 \big\{ \, \exp\left( 4 \epsilon \left[ \, f_\rho(1,3) 
 \right. \right. \nonumber \\
& & \hspace{0.6cm}   \left. \left. 
- f_\rho(1,4) - f_\rho(2,3) + f_\rho(2,4) \, \right] \, \right)- 1 \, \big\} 
\nonumber \\
& & 
+ \frac{1}{8} (y_{1/2}^2 + y_{2\rho}^2) \exp\left( -4 f_\rho(3,4) \right) 
 \big\{ \, \exp\left( 2 \epsilon \left[ \, f_\rho(1,3) 
 \right. \right. \nonumber \\
& & \hspace{0.3cm}  \left. \left. - f_\rho(1,4) - f_\rho(2,3) 
+ f_\rho(2,4) \, \right] \,  \right) - 1 \, \big\} 
\nonumber \\
& & 
+ \frac{1}{16} (y_{ps}^2 + y_{pc}^2) 
 \exp\left\{ \, - f_\rho(3,4) - f_\sigma(3,4) \, \right\} 
 \nonumber \\
& & \hspace{0.4cm} 
\times \big\{ \, \exp\left( \, \epsilon \left[ \, f_\rho(1,3) - f_\rho(1,4) 
 \right. \right. \nonumber \\
& & \hspace{0.4cm}  
 \left. \left. - f_\rho(2,3) + f_\rho(2,4) \, \right] \, \right) -1 \, 
              \big\} \, 
 \Big\} 
 \nonumber \\
&+& 
\sum_{\epsilon=\pm} 
         \int d^2{\tilde r}_3 d^2{\tilde r}_4 d^2{\tilde r}_5 \, 
 \Big\{ \, 
 \frac{1}{16}
  ( y_{1/4} y_{1/2}^2 - y_{1/4} y_{2\rho}^2)
\nonumber \\
& &
 \times \exp\left( \, - 8 f_\rho(3,4) 
  - 8 f_\rho(3,5) + 4 f_\rho(4,5) \, \right) 
\nonumber \\
& & 
 \times \big\{ \, \exp\left( \, 2 \epsilon \left[ \, 
        2 f_\rho(1,3) - f_\rho(1,4) 
        - f_\rho(1,5) \nonumber \right. \right. \\ 
& &
\left. \left.
        - 2 f_\rho(2,3) 
        + f_\rho(2,4) + f_\rho(2,5) \, \right] 
        \, \right) \, -1 \, \big\}
 \nonumber \\
&-&  
  \frac{1}{32} 
  (y_{2\rho} y_{ps}^2 - y_{2\rho} y_{pc}^2 -2 y_{1/2} y_{ps} y_{pc}) 
  \nonumber \\
& &
 \times \exp\left( \, - 2 f_\rho(3,4) - 2 f_\rho(3,5) 
   + f_\rho(4,5) - f_\sigma(4,5) \, \right) 
 \nonumber \\
& & 
 \times \big\{ \, \exp\left( \, \epsilon \left[ \, 
  2 f_\rho(1,3) - f_\rho(1,4) 
  - f_\rho(1,5) \right. \right. \nonumber \\
& & \left. \left.
- 2 f_\rho(2,3) 
  + f_\rho(2,4) + f_\rho(2,5) \, \right] \, \right)
  - 1 \, \big\}
\nonumber \\
&-& 
  \frac{1}{32} 
  (y_\sigma y_{ps}^2 + y_\sigma y_{pc}^2) \nonumber \\
& &
 \times \exp\left( \, - f_\rho(4,5) - 2 f_\sigma(3,4) 
   - 2 f_\sigma(3,5) + f_\sigma(4,5) \, \right)
 \nonumber \\
& &  
 \times \big\{ \, \exp\left( \, \epsilon \left[ \, 
 f_\rho(1,4) - f_\rho(1,5) \right. \right. \nonumber \\
& &
\left. \left. \hspace{0.3cm}
 - f_\rho(2,4) - f_\rho(2,5) \, \right] \, \right) - 1 \, \big\} \, 
 \Big\} \; \Bigg]
 \virg \nonumber \\
\label{eq:A2}
\end{eqnarray}
%==================================================================
where $f_{\rho(\sigma)}(i,j) 
       = K_{\rho(\sigma)} U(\vec{r}_i - \vec{r}_j)$. 
Noting the renormalization form, one obtain 
%=================(A3)=======================================
\begin{eqnarray}
y_{1/4}^{\rm{eff}} &=& y_{1/4}
+ \frac{1}{4} \, (y_{1/2}^2 - y_{2\rho}^2) \int_\alpha^\infty
\frac{d r}{\alpha} \left( \frac{r}{\alpha} \right)^{1 + 4 K_\rho } 
 \virg \label{eq:y_1/4^eff} 
 \nonumber \\
y_{1/2}^{\rm{eff}} &=& y_{1/2}
+ \frac{1}{2} \, y_{1/4} \, y_{1/2} \int_\alpha^\infty
\frac{d r}{\alpha} \left( \frac{r}{\alpha} \right)^{1 - 8 K_\rho } \nonumber \\
& & \hspace{0.7cm} + \frac{1}{4} \, y_{ps} y_{pc} \int_\alpha^\infty 
\frac{d r}{\alpha} \left( \frac{r}{\alpha} \right)^{1 + K_\rho - K_\sigma} 
 \virg \label{eq:y_1/2^eff} 
 \nonumber \\
y_{2\rho}^{\rm{eff}} &=& y_{2\rho}
- \frac{1}{2} \, y_{1/4} \, y_{2\rho} \int_\alpha^\infty
\frac{d r}{\alpha} \left( \frac{r}{\alpha} \right)^{1 - 8 K_\rho } \nonumber \\
& & \hspace{0.7cm} - \frac{1}{8} \, (y_{ps}^2 - y_{pc}^2) \int_\alpha^\infty 
\frac{d r}{\alpha} \left( \frac{r}{\alpha} \right)^{1 + K_\rho - K_\sigma} 
 \virg \label{eq:y_2t^eff} 
 \nonumber \\
y_{ps}^{\rm{eff}} &=& y_{ps}
+ \frac{1}{2} \, (y_{1/2} \, y_{pc} - y_{2\rho} y_{ps}) \, \int_\alpha^{\infty}
\frac{d r}{\alpha} \left( \frac{r}{\alpha} \right)^{1 - 2 K_\rho} \nonumber \\
& & \hspace{0.5cm} - \frac{1}{2} \, y_\sigma \, y_{ps} \int_\alpha^{\infty}
\frac{d r}{\alpha} \left( \frac{r}{\alpha} \right)^{1 - 2 K_\sigma}
\label{eq:y_d^eff} \virg  
\nonumber \\
y_{pc}^{\rm{eff}} &=& y_{pc}
+ \frac{1}{2} \, (y_{1/2} \, y_{ps} + y_{2\rho} y_{pc}) \, \int_\alpha^{\infty}
\frac{d r}{\alpha} \left( \frac{r}{\alpha} \right)^{1 - 2 K_\rho} \nonumber \\
& & \hspace{0.5cm} - \frac{1}{2} \, y_\sigma \, y_{pc} \int_\alpha^{\infty}
\frac{d r}{\alpha} \left( \frac{r}{\alpha} \right)^{1 - 2 K_\sigma}
\label{eq:y_d^eff} \point 
\label{eq:y^eff}
\end{eqnarray}
%==================================================================
By reexponentiating the renormalized form of eq.~(\ref{eq:A2}), 
 one obtain 
%=================(A4)============================================
\begin{eqnarray}
R(\vec{r}_1 - \vec{r}_2) &=& 
\exp\left( \, - K_\rho^{\rm{eff}} 
U(\vec{r}_1 - \vec{r}_2) \, \right)  \virg
\end{eqnarray}
%==================================================================
where
%================(A5)=========================================
\begin{eqnarray}
K_\rho^{\rm{eff}} &=&
K_\rho - \Bigg[ \, 2 \, y_{1/4}^2 
         \int_\alpha^\infty \frac{d r}{\alpha} 
         \left( \frac{r}{\alpha} \right)^{3 - 16 \, K_\rho}  
 \nonumber \\
& & \hspace{0cm}
- \, \frac{1}{2} \, (y_{1/2}^2 + y_{2\rho}^2) \, 
         \int_\alpha^\infty \frac{d r}{\alpha} 
         \left( \frac{r}{\alpha} \right)^{3 - 4 \, K_\rho}  
 \nonumber \\
& & \hspace{0cm}  
       - \, \frac{1}{16} \, (y_{ps}^2 + y_{pc}^2) \, 
         \int_\alpha^\infty \frac{d r}{\alpha} 
         \left( \frac{r}{\alpha} \right)^{3 - K_\rho - K_\sigma} \, 
 \Bigg] \, K_\rho^2 \point \nonumber \\
\label{eq:K^eff}
\end{eqnarray}
%=================================================================
From the condition of scale invariance for $\alpha \to \alpha(1 + dl)$,
 we obtain the RG equations of $K_\rho, K_\sigma, y_{1/4}, y_{1/2}, 
  y_{2 \rho}, y_{ps}$ and $y_{pc}$ in eqs. (\ref{eq:RG}).
RG equations for $K_{\sigma}$ and $y_\sigma$ are obtained 
 in a similar way from the response function, 
$R(\vec{r}_1 - \vec{r}_2)= \left \langle \, T_\tau e^{{\rm i} \phi_+(\vec{r}_1)}
                            e^{- {\rm i} \phi_+(\vec{r}_2)}\, \right \rangle_{H^{\rm{eff}}} 
$.

Next, we derive the RG equation of 
 response function \cite{Giamarchi-res} $R_s$ written as
%================(A6)========================================
\begin{eqnarray}
& & R_s (\vec{r}_1 - \vec{r}_2) \nonumber \\
& & = \left \langle \, T_\tau \sin\theta_+(\vec{r}_1)
                            \cos\phi_+(\vec{r}_1)
                            \sin\theta_+(\vec{r}_2)
                            \cos\phi_+(\vec{r}_2) \, 
                            \right \rangle_{H^{\rm{eff}}} 
 \point \nonumber \\
\end{eqnarray}
%=====================================================================
We calculate $R_s$ up to the second order in $H_1^{{\rm{eff}}}$.
In a way similar to RG equations, one obtain
%===================(A7)=============================================
\begin{eqnarray}
R_s (\vec{r}_1 - \vec{r}_2) \!
&=& \! \frac{1}{4} \exp\left( - f_\rho(1,2) \! - \! f_\sigma(1,2) \right)
 \! \times  \! {\tilde R_s}(\vec{r}_1 - \vec{r}_2) \virg \nonumber \\
 \label{eq:R_s}
\end{eqnarray}
%==================================================================
where
%================(A8)===========================================
\begin{eqnarray}
& & {\tilde R_s}(\vec{r}_1 - \vec{r}_2) = 1 - \frac{1}{2} \, \int_\alpha d^2{\tilde r}_3 
\nonumber \\
& &
\times  \, \Big\{ \, y_{2 \rho} \, 
\exp\left( \, 2 \left[ \, f_\rho(1,2) - f_\rho(1,3) - f_\rho(2,3) \, \right] \, \right) 
 \nonumber \\ 
& & \hspace{0.3cm}
+ \, y_{\sigma} \, 
 \exp\left( \, 2 \left[ \, f_\sigma(1,2) - f_\sigma(1,3) - f_\sigma(2,3) \, \right] \, \right) 
 \, \Big\}
\nonumber \\ 
& & \hspace{0.3cm}
+ \, \Bigg[ \, \left\{ \, 
 2 \, y_{1/4}^2 \, 
 \int_\alpha^\infty 
 \frac{dr}{\alpha} \left( \frac{r}{\alpha} \right)^{3 - 16 K_\rho} 
 \right.  \nonumber \\
& & \hspace{0.3cm}
 + \frac{1}{2} \, (y_{1/2}^2 + y_{2 \rho}^2) \, 
 \int_\alpha^\infty 
 \frac{dr}{\alpha} \left( \frac{r}{\alpha} \right)^{3 - 4 K_\rho} 
 \nonumber \\ 
& & \nonumber \\ 
& & \hspace{0.3cm} 
\left. 
 + \frac{1}{16} \, (y_{ps}^2 + y_{pc}^2) \, \int_\alpha^\infty 
 \frac{dr}{\alpha} \left( \frac{r}{\alpha} \right)^{3 - K_\rho - K_\sigma}
 \right\} K_\rho^2 \nonumber \\
& & \nonumber \\ 
& & \hspace{0.3cm}
+ \, \left\{ \, 
 \frac{1}{2} \, y_{\sigma}^2 \, 
 \int_\alpha^\infty 
 \frac{dr}{\alpha} \left( \frac{r}{\alpha} \right)^{3 - 4 K_\sigma} 
 \right.
\nonumber \\ 
& & \nonumber \\ 
& & \hspace{0.3cm} 
\left. 
 + \frac{1}{16} \, (y_{ps}^2 + y_{pc}^2) \, 
 \int_\alpha^\infty 
 \frac{dr}{\alpha} \left( \frac{r}{\alpha} \right)^{3 - K_\rho - K_\sigma}
 \, \right\} K_\sigma^2 \, \Bigg] \nonumber \\
& & \nonumber \\ 
& & \hspace{0.5cm} 
 \times \, U({\vec r}_1 - {\vec r}_2) \point 
 \nonumber \\
\label{eq:tildeR_s}
\end{eqnarray}
%==================================================================
The scale transformation, 
 $\alpha \to \alpha^\prime = \alpha (1+dl)$, leads to 
$(
r = |\vec{r}_1-\vec{r}_2|=\alpha {\rm e}^l
)$
%=================(A9)========================================
\begin{eqnarray}
{\tilde R_s}(r,\alpha) &=& I(dl) \, {\tilde R_s}(r,\alpha^\prime) \virg
 \label{eq:multiplicative scale}
\end{eqnarray}
%==================================================================
where
%=================(A10)========================================
\begin{eqnarray}
& & I(dl) = \exp \Biggl[ \,
- y_{2 \rho} dl - y_\sigma dl 
\nonumber \\ 
& & \hspace{0cm}
+ \left\{ \, \left( 2 y_{1/4}^2 + \frac{1}{2} (y_{1/2}^2 + y_{2 \rho}^2) 
+ \frac{1}{16} (y_{ps}^2 + y_{pc}^2) \right) 
 K_\rho^2 \right.
\nonumber \\ 
& & \nonumber \\ 
& &  \hspace{0cm}
\left. 
+ \left( \frac{1}{2} y_{\sigma}^2 
+ \frac{1}{16} (y_{ps}^2 + y_{pc}^2) \right) 
 K_\sigma^2 \, \right\} U(\vec{r}_1 - \vec{r}_2) dl \; \Biggr] \point 
 \label{eq:I(dl)} \nonumber \\
\end{eqnarray}
%===================(A11)==================================
From eqs. (\ref{eq:R_s}), (\ref{eq:multiplicative scale}) and (\ref{eq:RG}), 
we obtain
%======================================================================
\begin{eqnarray}
& & R_s(r) = \nonumber \\
& & \frac{1}{4} \exp\left[ - \int_0^{\ln(r / \alpha )} \hspace{-0.4cm} dl 
\left\{ K_\rho(l) + K_\sigma(l) + y_{2 \rho}(l) + y_\sigma(l) \right\} 
 \right]
 \virg \nonumber \\
\end{eqnarray}
%==================================================================
 which leads to 
%--------------------------------------------------------------
\begin{eqnarray}
 \frac{d}{dl}\:R_s(l) \, &=& -\, [ \, K_\rho(l) 
 + K_\sigma(l) + y_{2\rho}(l) + y_\sigma(l) \, ] \, R_s(l) \point
  \nonumber \\
\end{eqnarray}
%==================================================================
We note that the additional term coupled to $R_{sc}$ can be calculated 
 in terms of the operator product expansion.
\cite{Cardy}

%-------------  References              ----

\end{document}